\documentclass{article}
\pdfoutput=1
\usepackage{microtype}
\usepackage{graphicx}
\usepackage{booktabs} % for professional tables

\usepackage{hyperref}

\usepackage[accepted]{icml2024}

\usepackage{amsmath}
\usepackage{amssymb}
\usepackage{mathtools}
\usepackage{amsthm}
\usepackage{hyperref}
\usepackage{url}
\usepackage{booktabs}       % professional-quality tables
\usepackage{amsfonts}       % blackboard math symbols
\usepackage{graphicx}
\usepackage{array}
\usepackage{multirow}
\usepackage{wrapfig}
\usepackage{algorithmic}
\usepackage{algorithm}
\usepackage{verbatim}
\usepackage{subcaption}
\usepackage{enumitem}
\usepackage{soul}
\usepackage{float}
\usepackage{xcolor}

\usepackage{amsmath,amsfonts,bm}

\def\eqref#1{equation~\ref{#1}}
\def\1{\bm{1}}

\def\rvx{{\mathbf{x}}}
\def\rvy{{\mathbf{y}}}

\DeclareMathAlphabet{\mathsfit}{\encodingdefault}{\sfdefault}{m}{sl}
\SetMathAlphabet{\mathsfit}{bold}{\encodingdefault}{\sfdefault}{bx}{n}

\theoremstyle{plain}

\theoremstyle{definition}

\theoremstyle{remark}

\usepackage[textsize=tiny]{todonotes}

\newcommand{\mi}{\mathbb{I}}

\newcommand{\p}{\mathrm{p}}
\newcommand{\q}{\mathrm{q}}
\newcommand{\ept}{\mathbb{E}}
\newcommand{\ent}{\mathbb{H}}
\newcommand{\kl}{D_{\mathrm{KL}}}

\newcommand{\ds}{\mathcal{D}}
\newcommand{\J}{\mathcal{J}^{\mathrm{info}}}

\definecolor{asparagus}{rgb}{0.53, 0.66, 0.42}
\definecolor{ballblue}{rgb}{0.13, 0.67, 0.8}

\icmltitlerunning{InfoNet: Neural Estimation of Mutual Information without Test-Time Optimization}

\begin{document}

\twocolumn[
\icmltitle{InfoNet: Neural Estimation of Mutual Information \\
without Test-Time Optimization}



\icmlsetsymbol{equal}{*}

\begin{icmlauthorlist}
\icmlauthor{Zhengyang Hu}{hku}
\icmlauthor{Song Kang}{ustc}
\icmlauthor{Qunsong Zeng}{hku}
\icmlauthor{Kaibin Huang}{hku}
\icmlauthor{Yanchao Yang}{hku}
\end{icmlauthorlist}

\icmlaffiliation{hku}{The Department of Electrical and Electronic Engineering, the University of Hong Kong}
\icmlaffiliation{ustc}{University of Science and Technology of China}

\icmlcorrespondingauthor{Yanchao Yang}{yanchaoy@hku.hk}

\icmlkeywords{Mutual Information, Neural Estimation}

\vskip 0.3in
]



\printAffiliationsAndNotice{}  % leave blank if no need to mention equal contribution

\begin{abstract}
Estimating mutual correlations between random variables or data streams is essential for intelligent behavior and decision-making. 
As a fundamental quantity for measuring statistical relationships, mutual information has been extensively studied and utilized for its generality and equitability. 
However, existing methods often lack the {\it efficiency} needed for real-time applications, such as test-time optimization of a neural network, or the {\it differentiability} required for end-to-end learning, like histograms.
We introduce a neural network called {\it InfoNet}, which directly outputs mutual information estimations of data streams by leveraging the attention mechanism and the computational efficiency of deep learning infrastructures. 
By maximizing a dual formulation of mutual information through {\it large-scale simulated training}, our approach circumvents time-consuming test-time optimization and offers generalization ability.
We evaluate the {\it effectiveness} and {\it generalization} of our proposed mutual information estimation scheme on various families of distributions and applications. 
Our results demonstrate that InfoNet and its training process provide a graceful {\it efficiency-accuracy} trade-off and {\it order-preserving} properties. 
We will make the code and models available as a comprehensive toolbox to facilitate studies in different fields requiring real-time mutual information estimation.
\end{abstract}

\section{Introduction}\label{sec:intro}
\vspace{-0mm}

We exist in a universe where various entities are interconnected. 
At the micro level, particles can exhibit entanglement, as described by quantum mechanics, 
while at the macro level, celestial bodies are governed by gravity, characterized by general relativity. 
These interconnections ensure that our observations of the states of different entities around us are intricately correlated rather than independently distributed. 
This interconnectedness enables us to make informed reasoning and predictions.

\begin{figure}[!t]
    \centering
    \vspace{-10pt}
    \includegraphics[width=0.7\linewidth]{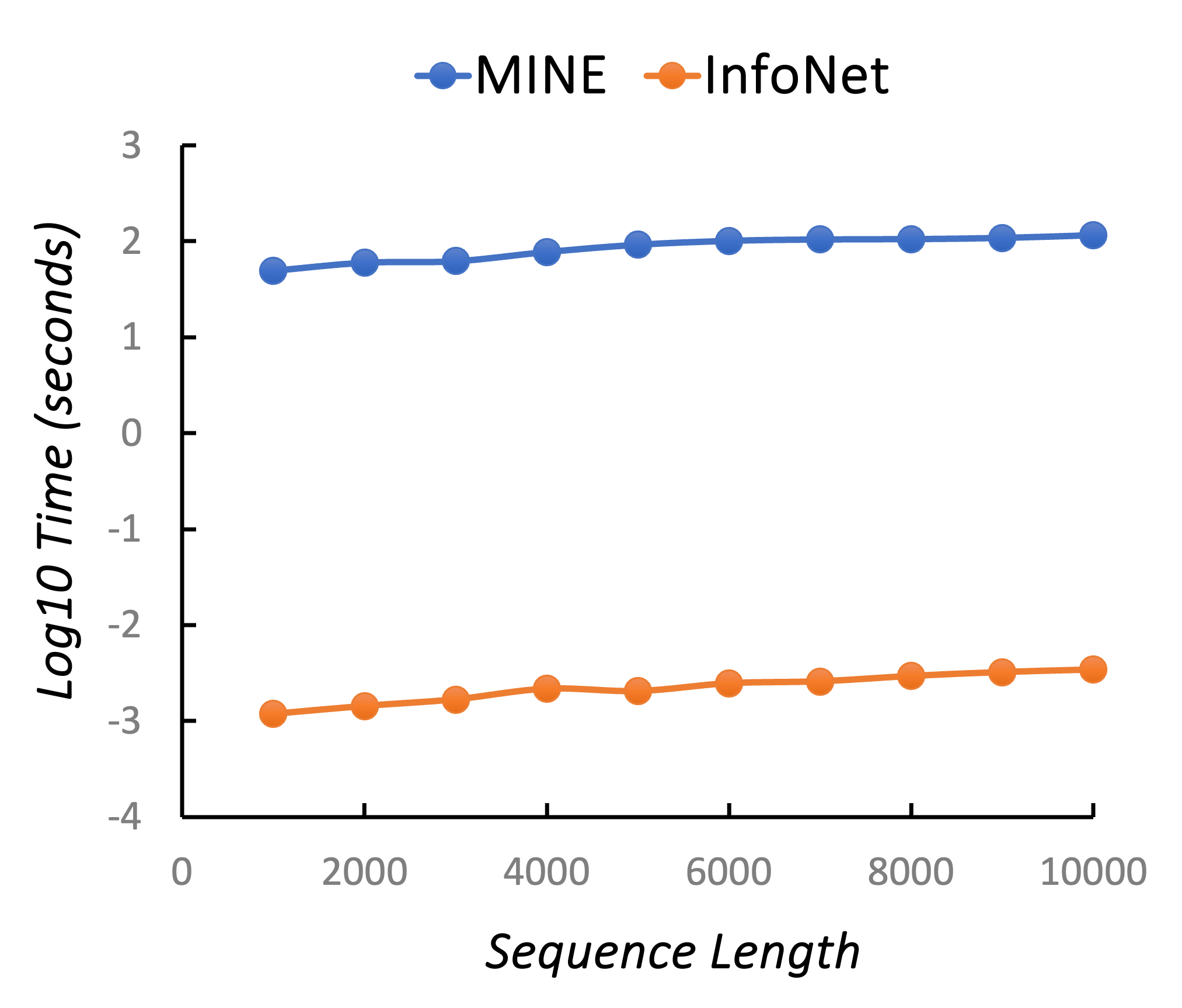}
    \vspace{-10pt}
    \caption{Log-scale run time comparison of MINE \citep{MINE} and the proposed InfoNet, which consistently achieves faster performance by magnitudes across sequences of varying lengths by bypassing the costly test-time optimization.}
    \label{fig:efficiency-accuracy-tradeoff}
    \vspace{-20pt}
\end{figure}

\begin{figure*}[!ht]
    \centering
    \includegraphics[width=0.92\textwidth]{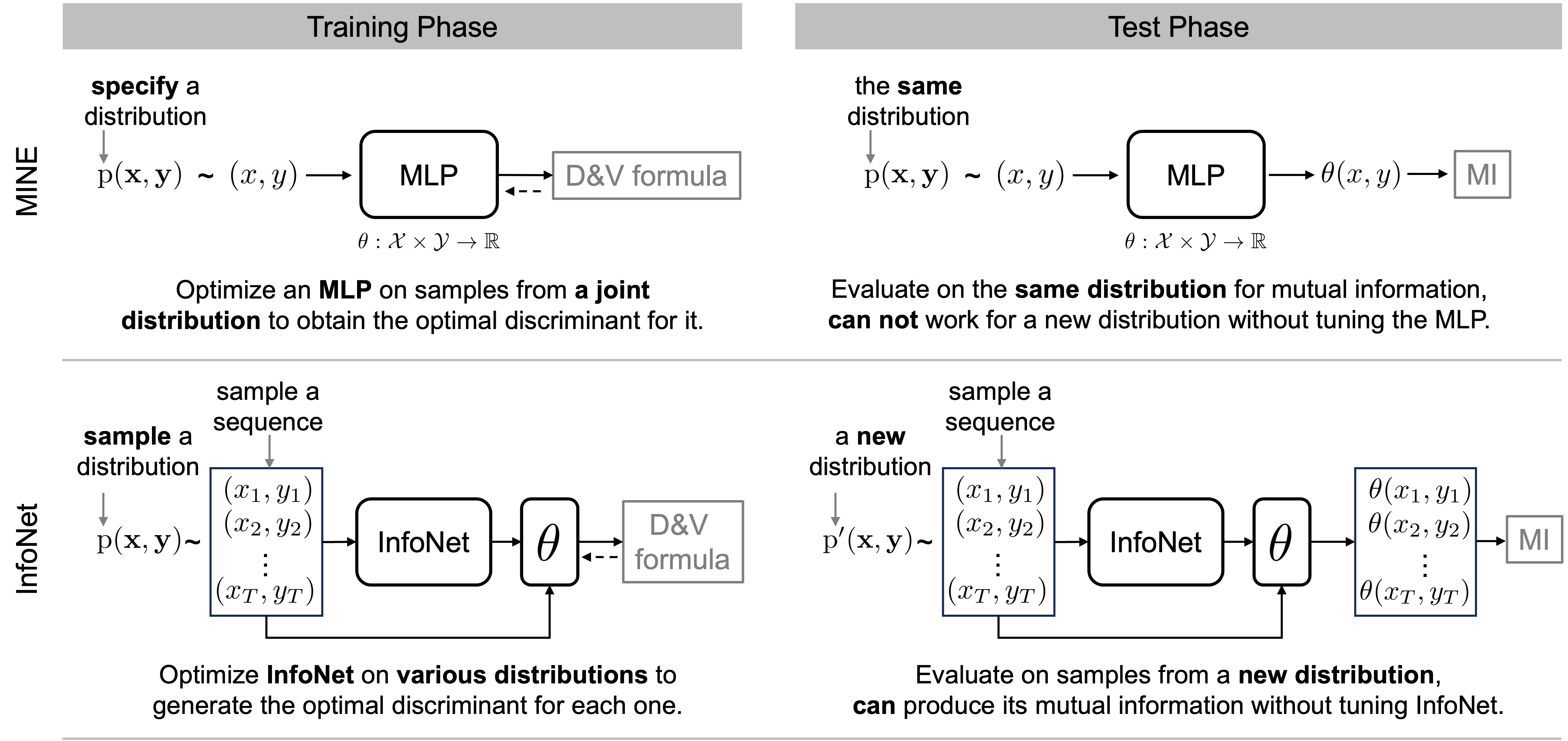}
    \vspace{-1mm}
    \caption{A comparison of MINE \citep{MINE} and the proposed InfoNet for neural MI estimation. In the training phase, MINE optimizes an MLP's parameters (as a discriminant function) using the dual formula \citep{donsker1983asymptotic} against a joint distribution. The optimized MLP then estimates the same distribution's MI with its samples. However, the MLP is not optimal for a new distribution and requires retraining (test-time optimization) before providing an estimate. In contrast, InfoNet is trained on various distributions to output the optimal discriminant ($\theta$) for any distribution. At test time, InfoNet predicts the optimal discriminant for a new distribution using its samples, leveraging the generalization capability from large-scale training, thus eliminating the need for test-time optimization and increasing efficiency.}
    \vspace{-3mm}
    \label{fig:mine-vs-infonet} 
\end{figure*}

Efficiently estimating correlations between scene entities from environmental sensory signals is essential for the emergence of intelligent behavior. 
This is particularly relevant for embodied agents that interact with the scene and receive large volumes of streaming data, such as video, audio, and touch, within seconds. 
Rapid correlation estimation helps agents build informative representations of their surroundings and identify crucial elements for survival.
Moreover, vast amounts of data are generated every second across the internet, including stock prices, social media messages, e-commerce transactions, and Internet-of-Things devices. 
Efficiently estimating mutual correlations between different types or parts of this data informs critical analyses for decision-making. 

In this work, we study how to {\it neuralize} the computation of mutual information (MI) between two random variables from sequences sampled from their {\it empirical} joint distribution. 
Specifically, we want to explore whether the estimation of MI can be performed by a neural network without test-time optimization, i.e., taking a pair of sequences as input and speeding out the MI estimate without re-training the network, which guarantees efficiency and differentiability of the estimation procedure. 

As a fundamental concept in information theory \citep{shannon1948mathematical}, a huge amount of effort has been devoted to the estimation of MI \citep{paninski2003estimation,kraskov2004estimating}, due to its generality and equitability \citep{reshef2011detecting,kinney2014equitability}.
For example, many algorithms have been proposed to improve the accuracy and efficiency of MI estimation, which include both non-parametric  
and parametric methods. 
However, most of them do not utilize neural networks and can not benefit from advances in deep learning techniques.
Recently, MINE \citep{MINE} employs a dual formulation of the Kullback–Leibler divergence and estimates the MI of a pair of sequences by optimizing a neural network's parameters against the dual objective. 
Even though the estimation can be performed via back-propagation, the optimization process is still behind real-time (Fig.~\ref{fig:efficiency-accuracy-tradeoff}, where a joint sequence is sampled from a randomly generated mixture of Gaussian). 
Moreover, each time the joint distribution changes, a new optimization has to be performed (e.g., the network in MINE is only optimized for a specific distribution, also see Fig.~\ref{fig:mine-vs-infonet}, first row), thus not efficient.

To overcome these difficulties, yet still enjoy the {\it efficiency} of deep networks, 
we propose a novel network architecture that leverages the attention mechanism \citep{vaswani2017attention} and encodes the aforementioned optimization into the network parameters.
Specifically, the proposed network takes as input a sequence of observations (pairs) and outputs a tensor, which aims at maximizing the Donsker-Varadhan \citep{donsker1983asymptotic} dual and can be converted into an MI estimate by a quick summation over different entries.
This way, we {\it transform} the optimization-based estimation into a feed-forward prediction, thus {\it bypassing} the time-consuming test-time gradient computation and avoiding sub-optimality via large-scale training on a {\it wide spectrum} of distributions.
Our experiments demonstrate efficiency, accuracy and generalization of the proposed MI neural estimation framework.

In summary, we: 1) propose a neural network and training method for efficiently estimating MI of any distribution (sequences) without resorting to test-time optimization; 
2) conduct an extensive study on the proposed scheme's effectiveness with different distribution families, verifying its accuracy and order-preserving properties; 
and 3) demonstrate the generalization of the proposed InfoNet on real-world distributions, showcasing promising results in object discovery from videos.

\vspace{-2mm}
\section{Problem Statement}\label{sec:prob-state}

We consider real-world scenarios where an agent receives sensory inputs via multiple channels, i.e., multimodal signals.
We treat these observations as random variables and their (synchronized) temporal sequences as if sampled from an empirical joint distribution. 
More explicitly, we characterize observations $\{(x_t,y_t)\}_{t=1}^T$ as samples from a joint distribution $\p(\rvx,\rvy)$, e.g., by histogramming.
Our goal is to compute Shannon's MI between $\rvx$ and $\rvy$, i.e.,  $\mi(\rvx,\rvy)$, in an efficient manner such that an agent can leverage these correlations to learn useful representations and to make effective decisions.
Specifically, we aim to train neural networks $\phi$ such that $\mathcal{C}(\rvx,\rvy)=\phi(\{(x_t,y_t)\})$ is an estimation of the MI of $\p(\rvx,\rvy)$ from the input sequences, without re-training $\phi$ for different distributions (see Fig.~\ref{fig:mine-vs-infonet}, second row).
In this work, we focus on the efficient computation of low-dimensional random variables, e.g., 1D/2D, and leverage the projection technique in \citet{goldfeld2021sliced} for an extension to high-dimensional while maintaining computational efficiency and accuracy.

\vspace{-1mm}
\section{Neural MI Estimation without Test-Time Optimization}\label{sec:method}

\begin{figure*}[!t]
    \centering
    \includegraphics[width=0.75\textwidth]{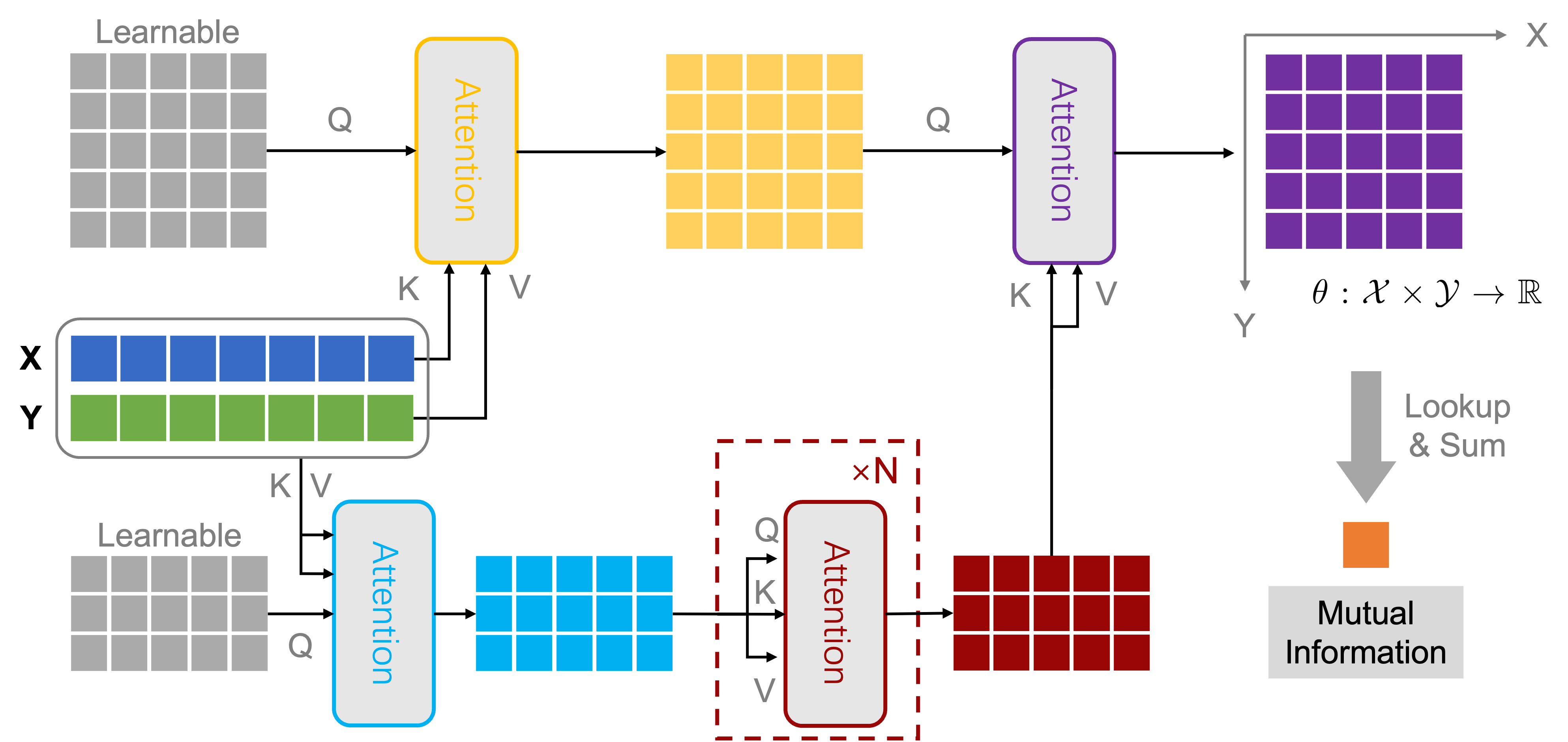}
    \vspace{-1mm}
    \caption{The proposed InfoNet architecture for MI prediction comprises learnable queries and attention blocks. It accepts a sequence of samples from two random variables and outputs a look-up table (top-right) representing a discretization of the optimal scalar discriminant function defined on the joint domain in the Donsker-Varadhan representation \citep{donsker1983asymptotic}. 
    The MI between the two random variables (sequences) can then be calculated by summation according to Eq.~\ref{eq:mi-donvara}.
    Note that the input sequences for training are sampled from various distributions. Please also refer to Fig.~\ref{fig:mine-vs-infonet} for a comparison between MINE and InfoNet training schemes. 
    }
    \vspace{-3mm}
    \label{fig:architectures-mi} 
\end{figure*}

MI can be written in Shannon Entropy: $\mi(\rvx,\rvy)=\ent(\rvx)-\ent(\rvx|\rvy)$, or in Kullback–Leibler divergence: $\mi(\rvx,\rvy)=\kl(\p_{\rvx,\rvy}\|\p_{\rvx}\cdot\p_{\rvy})$. However, exact computation is only feasible for discrete variables or a restricted set of distributions \citep{paninski2003estimation}.
Recently, MINE \citep{MINE} proposes estimating MI using a neural network trained with a dual formula \citep{donsker1983asymptotic}. 
This method is capable of handling continuous random variables, but requires training from scratch for a different joint distributions $\p'(\rvx,\rvy)$ (test-time optimization), making real-time MI estimation challenging.

In the following, we provide details of the dual formulation \citep{donsker1983asymptotic} employed for MI estimation and elaborate on the proposed methods for training the neural network $\phi$ for computing MI of an unseen distribution without test-time optimization.

\paragraph{Dual Estimation of MI} 
According to \citet{donsker1983asymptotic}
(also see \citet{gutmann2010noise}), 
the KL-divergence between two distributions, $p$ and $q$, can be written as: $\kl(\p\|\q)=\sup_{\theta}\ept_{\p}[\theta]-\log(\ept_{\q}[\exp(\theta)])$, where $\theta$ is a discriminant function, whose output is a scalar value, defined on the joint domain with finite expectations. 
The dual estimation formula for MI is then as follows:
\begin{align}
    \mi(\rvx,\rvy)
    &= \sup_{\theta} \J(\theta; \rvx,\rvy) \nonumber\\
    &= \sup_{\theta} \ept_{\p_{\rvx,\rvy}}[\theta]-\log(\ept_{\p_{\rvx}\cdot\p_{\rvy}}[\exp(\theta)]),
    \label{eq:mi-donvara}
\end{align}
with $\theta: \mathcal{X} \times \mathcal{Y} \rightarrow \mathbb{R}$ and $\mathcal{X},\mathcal{Y}$ the domain of the random variables $\rvx,\rvy$ correlated by a joint distribution $\p(\rvx,\rvy)$.
One can instantiate $\theta$ as a neural network and train it with the right-hand side in Eq.~\ref{eq:mi-donvara} as the objective, as done in MINE \citep{MINE}.
The optimal value of the right hand can then serve as the estimate of MI between $\rvx$ and $\rvy$ under $\p(\rvx,\rvy)$.
The same training has to be performed for a new distribution, i.e., test-time optimization (see Fig.~\ref{fig:mine-vs-infonet}, first row).
In contrast, we propose to bypass the test-time optimization by training a novel network architecture that directly outputs the optimal discriminant regarding the dual, using samples from the new distribution.
In other words,
we treat the {\it optimal} scalar-valued function $\theta$ of a new distribution as the output of the neural network $\phi$. 
This way, we can speed up the estimation by magnitudes and enjoy the benefit of the differentiability of deep neural networks.

\vspace{-1mm}
\paragraph{Optimal Discriminant Prediction}
To enable predicting the optimal discriminant $\theta$ of a distribution $\p(\rvx,\rvy)$ from $\p$'s samples $\{(x_t,y_t)\}$, 
we formalize $\theta_{\rvx,\rvy}=\phi(\{(x_t,y_t)\})\in\mathbb{R}^{L\times L}$ as a 2D tensor, 
where $L$ represents the quantization levels of the range of the involved random variables.
Now, the value of $\theta_{\rvx,\rvy}(x_t,y_t)$ for a continuous pair $(x_t,y_t)$ can be directly read out from the tensor as a look-up table with correct indexing and appropriate interpolation.

To facilitate the prediction, we design a neural network by adapting the attention mechanism described in \citep{jaegle2021perceiver}.
The proposed network structure $\phi$ is illustrated in Fig.~\ref{fig:architectures-mi} and named as {\it InfoNet}.
It takes in a pair of jointly sampled sequences, e.g., $\{(x_t,y_t)\}_{t=1}^T$, and outputs a tensor $\theta_{\rvx,\rvy}$ as the discretization of the scalar function $\theta$ in Eq.~\ref{eq:mi-donvara}.

In addition, to improve the training efficiency,
we apply a copula transformation \citet{durante2010copula} on the sequences before inputting them to the network, which helps normalize their range to $[0,1]$. 
It is worth noting that the invariance property of MI under bijective mappings of the RVs ensures that such a transformation does not change the MI between the two sequences. 
More details on this copula transformation can be found in Appendix~\ref{sec:intro-copula}.

With the predicted (discretized) discriminant function $\theta_{\rvx,\rvy}$, we can then compute an estimate of the MI between $\rvx$ and $\rvy$ using the quantity $\J(\theta; \rvx,\rvy)$ in Eq.~\ref{eq:mi-donvara}.

To ensure that the predicted discriminant is optimal for $\p(\rvx,\rvy)$ under Eq.~\ref{eq:mi-donvara},
we train the neural network $\phi$ using the following objective (a discretization of Eq.~\ref{eq:mi-donvara}):
\begin{align}
    \mathcal{L}_{\text{MI}}(\phi,\mathcal{D}) 
    &= \frac{1}{N}\sum_{i=1}^N\mathcal{J}(\theta_{\mathbf{x}^i,\mathbf{y}^i};\mathbf{x}^i,\mathbf{y}^i)\nonumber\\
    &= \frac{1}{N}\sum_{i=1}^N\Bigg\{
    \frac{1}{T}\sum_{t=1}^T\theta_{\mathbf{x}^i,\mathbf{y}^i}(x_t^i,y_t^i)\nonumber \\
    &\quad -\log\left(\frac{1}{T}\sum_{t=1}^T\exp(\theta_{\mathbf{x}^i,\mathbf{y}^i}(x_t^i,\tilde{y}_t^i))\right)
    \Bigg\}.
    \label{eq:explict-mi-loss}
\end{align}
Here $\ds$ is a dataset of $N$ {\it different distributions}, i.e., $\ds=\{(\rvx^i,\rvy^i)\}_{i=1}^N$ with each $(\rvx^i,\rvy^i)=\{(x^i_t,y^i_t)\}_{t=1}^T$ representing a sequence sampled from a distribution $\p^i(\rvx,\rvy)$. 
Please also note that the second expectation in Eq.~\ref{eq:mi-donvara} is over the product of the marginals, so we write 
$\tilde{y}_i$ in the second summation in Eq.~\ref{eq:explict-mi-loss} to emphasize the difference.
The marginal distribution $\p^i(\rvy)$ can be sampled by simply breaking the pairing between $x^i_t$ and $y^i_t$, e.g., by shuffling $\{y^i_t\}_{t=1}^T$ as if $x$ does not exist.
We detail the generation of the training samples of many different distributions in Sec.~\ref{sec:datagen}. 
Since the training of $\phi$ (InfoNet) is performed with a large set of (simulated) distributions between $\rvx$ and $\rvy$ instead of a single distribution as in MINE \citep{MINE}, 
the predicted $\theta_{\rvx',\rvy'}$ from samples 
$\{(x'_t,y'_t)\}_{t=1}^T$ of a new distribution $\p'(\rvx,\rvy)$
is supposed to maximize the quantity $\J$ in Eq.~\ref{eq:mi-donvara} through generalization. 
We also verify the generalization of the trained InfoNet with an extensive study in the experimental section.

\vspace{-2mm}
\section{Data Generation and Training Algorithm}
\label{sec:datagen}

\begin{algorithm}[!h]
   \caption{InfoNet Training}\label{alg:train} 
\begin{algorithmic}[1] 
    \REQUIRE A maximum number of Gaussian components; Learning rate $\eta$\\
    \REPEAT
    \STATE Randomly select $N$ two-dimensional Gaussian mixture distributions
    \STATE Select $T$ data points from each joint distribution
    \STATE Shuffle the $\rvy$ component to get its marginal samples
    \STATE Put joint samples into the model and get $N$ two-dimensional lookup tables $\theta_{\rvx^i,\rvy^i}$'s
    \STATE Apply lookup function to get the corresponding discriminant values $\theta_{\rvx^i,\rvy^i}(x_t^i,y_t^i)$ for all data points in the joint and marginal samples
    \STATE $\begin{aligned}
\mathcal{L} &\leftarrow \frac{1}{N}\sum_{i=1}^N\left\{
    \frac{1}{T}\sum_{t=1}^T\theta_{\mathbf{x}^i,\mathbf{y}^i}(x_t^i,y_t^i) \right.\\
    &\quad \left.-\log\left(\frac{1}{T}\sum_{t=1}^T\exp(\theta_{\mathbf{x}^i,\mathbf{y}^i}(x_t^i,\tilde{y}_t^i))\right)
    \right\}.
\end{aligned}$
    \STATE Do gradient ascent for $\mathcal{L}$\\
    \UNTIL \textbf{convergence.}  
\end{algorithmic}
\end{algorithm}

To generate training data, 
we consider sampling the joint distributions (sequences) 
$\ds=\{(\rvx^i,\rvy^i)\}_{i=1}^N$
from Gaussian Mixture Models (GMMs).
It is widely accepted that GMM is a versatile and effective tool for modeling real-world distributions due to its capability to handle complex and noisy data \cite{reynolds2009gaussian}.
Specifically,
GMMs represent a family of distributions as a weighted sum of Gaussian components defined as: $\p(z)=\sum_{i=1}^{K}\pi_i\mathcal{N}(z|\mu_i, \Sigma_i),$ where $\p(z)$ is the probability density function (PDF) of the GMM, $K$ is the total number of components in the mixture, $\pi_i$ denotes the weight of the $i$-th component satisfying $\sum_{i=1}^K\pi_i=1$, and $\mathcal{N}(z|\mu_i, \Sigma_i)$ is the PDF of a Gaussian with mean $\mu_i$ and covariance $\Sigma_i$.
By varying the parameters $K$, $\pi_i$, $\mu_i$, and $\Sigma_i$, a GMM can faithfully approximate an arbitrary distribution.
With this, we propose that sampling from GMMs 
allows us to synthesize arbitrarily complex distributions so that the trained InfoNet can generalize to real-world ones (in a similar spirit to \citet{cranmer2020frontier,lavin2021simulation}).

In our experiments, we set the maximum number of components to 20 to ensure enough diversity in the sampled GMMs.
Specifically, we first randomly choose a number $K$ from $\{1,2,...,20\}$, and then perform another sampling of the component weights $\{\pi_i\}_{i=1}^K$ such that their sum is one.
For each GMM component, we randomly sample its mean from the interval $[-5, 5]$. 
To generate the covariance matrix, we begin by creating a matrix $\mathbf{D}$ where each element is sampled from the range $[-3, 3]$. Then, the covariance matrix is derived by $\Sigma = \mathbf{D}\mathbf{D}^T + \epsilon\mathbf{I}$, where $\epsilon = 0.01$ is to enforce the matrix to be positive definite.
To this end, a random GMM distribution is instantiated, and we can sample from it to get a joint sequence by partitioning $z$ into two parts.
Examples of the GMM samplings can be found in Appendix~\ref{sec:data-distribution}.
A training batch contains 32 randomly generated GMM distributions (sequences) with a sample length of 2000. 
Also, note that each batch is sampled from a different set of GMMs to ensure the training data for InfoNet is diverse and can explore the whole GMM family. 
Trained with randomly sampled distributions, 
our model is empowered to estimate MI for an untrained distribution encountered during test time.
Please refer to Algorithm~\ref{alg:train} for the full training pipeline of the proposed method.

\vspace{-2mm}
\section{Experiments}\label{sec:experiments}

We concentrate on three aspects related to the training effectiveness and estimation efficiency: 
1) Establishing evaluation criteria and collecting data for the proposed InfoNet and baseline methods;
2) Validating and comparing the proposed MI estimation pipeline with other baselines in various settings; 
3) Conducting experiments on data with real-world statistics to evaluate performance against other baselines in terms of efficiency and generalization.

\vspace{-2mm}
\subsection{Evaluation Data and Metrics}\label{sec:eval-method}

\paragraph{Evaluation} sequences (distributions) are generated with the same protocol described in Sec.~\ref{sec:datagen} 
and the ground-truth (GT) MI is determined by the following: 
for a single-component GMM (Gaussian), 
we apply the analytical formula of MI; otherwise,
the Monte-Carlo Integration (MCI) method \citep{shapiro2003monte} is employed to compute the GT by integrating the known density function.

\vspace{-2mm}
\paragraph{Setups and Metrics.}\label{para:setup}
We evaluate our method and others with the following setups:
\vspace{-3mm}
\begin{itemize}[leftmargin=*]

    \item {\bf Test-Time Efficiency.} 
    We compare the computational efficiency 
    of the proposed InfoNet with various baseline methods across different distributions and sequence lengths drawn from the GMM family.
    \vspace{-2.2mm}
    \item {\bf Sanity Check.}
    We use the sequences sampled from Gaussian distributions to benchmark against other methods, a commonly adopted evaluation setting in the MI estimation literature \citep{belghazi2018mutual,piras2023robust}. The GT MI values are computed with the analytical formula.
    \vspace{-2.2mm}
    \item {\bf GMMs with Multiple Components.} 
    We also analyze the mean and variance of the errors in estimated MI. Specifically, we bin the sampled evaluation sequences based on their ground-truth MI values, such as those with a ground-truth MI of around 0.5. We then report the mean and variance of these errors for each bin of different methods.
    \vspace{-6mm}
    \item {\bf Mutual Correlation Order Accuracy.}
    Beyond application domains where the exact MI value is critical, most of the time, 
    for decision-making, the more important is the order of mutual correlations between different random variables. For this, we generate an evaluation set of joint distributions consisting of triplets of random variables $\{(\rvx,\rvy,\rvy')\}$, whose ground-truth order is determined by the computed GT MI (i.e., $\mathbb{I}(\rvx,\rvy)>\mathbb{I}(\rvx,\rvy')$). We test different methods on the triplets to check the correlation order accuracy averaged over all triplets.
    \vspace{-2mm}
    \item {\bf High Dimensional Independence Testing.} 
    We further employ the slicing technique proposed in sliced mutual information (SMI) \citep{goldfeld2021sliced} for estimating MI between high-dimensional variables with InfoNet. Details are in Appendix~\ref{sec:intro-slicemi}. 
    Note that the slicing technique causes minor computational overhead due to parallelization.
    Due to the lack of GT values, we assess InfoNet's capability in accurately determining the independence between two random vectors.
    \vspace{-2mm}
    \item {\bf Evaluation with Motion Data.} 
    We verify the generalization 
    of the trained InfoNet on motion data with real-world statistics (e.g., \citep{radford2021learning,zheng2023pointodyssey}), where the goal is to check whether the points coming from the same object in motion can be grouped correctly by the estimated MI. 
\end{itemize}

\vspace{-4mm}
\subsection{Results and Comparisons}

\begin{table}[!t]
\setlength{\tabcolsep}{5pt}
\begin{footnotesize}
\caption{Comparison of test-time efficiency on GMM distributions with varying lengths (unit: seconds).}
\vspace{-3mm}
\label{tab:time_results}
\begin{center}
\begin{sc}
\begin{tabular}{c c c c c c c }
\toprule
 Seq. Length & 200 & 500 & 1000 & 2000 & 5000   \\
 
\midrule
KSG-1 & 0.009 & 0.024 & 0.049 & 0.098 & 0.249  \\
KSG-5 & 0.010 & 0.025 & 0.049 & 0.102 & 0.253  \\
KDE & 0.004 & 0.021 & 0.083 & 0.32 & 1.801  \\
\midrule
MINE-2000 & 3.350 & 3.455 & 3.607 & 3.930 & 4.157  \\ 
MINE-500 & 0.821 & 0.864 & 0.908 & 0.991 & 1.235   \\
MINE-10 & 0.017 & 0.017 & 0.019 & 0.021 & 0.027  \\
Ours-1 & 0.010 & 0.010 & 0.011 & 0.011 & 0.013  \\
Ours-16 & \textbf{0.001} & \textbf{0.002} & \textbf{0.002} & \textbf{0.002} & \textbf{0.003} \\

\bottomrule
\end{tabular}
\end{sc}
%\end{small}
\end{center}
\end{footnotesize}
\vspace{-5mm}
\end{table}

\begin{figure}[t]
    \centering
    \includegraphics[width=0.75\linewidth]{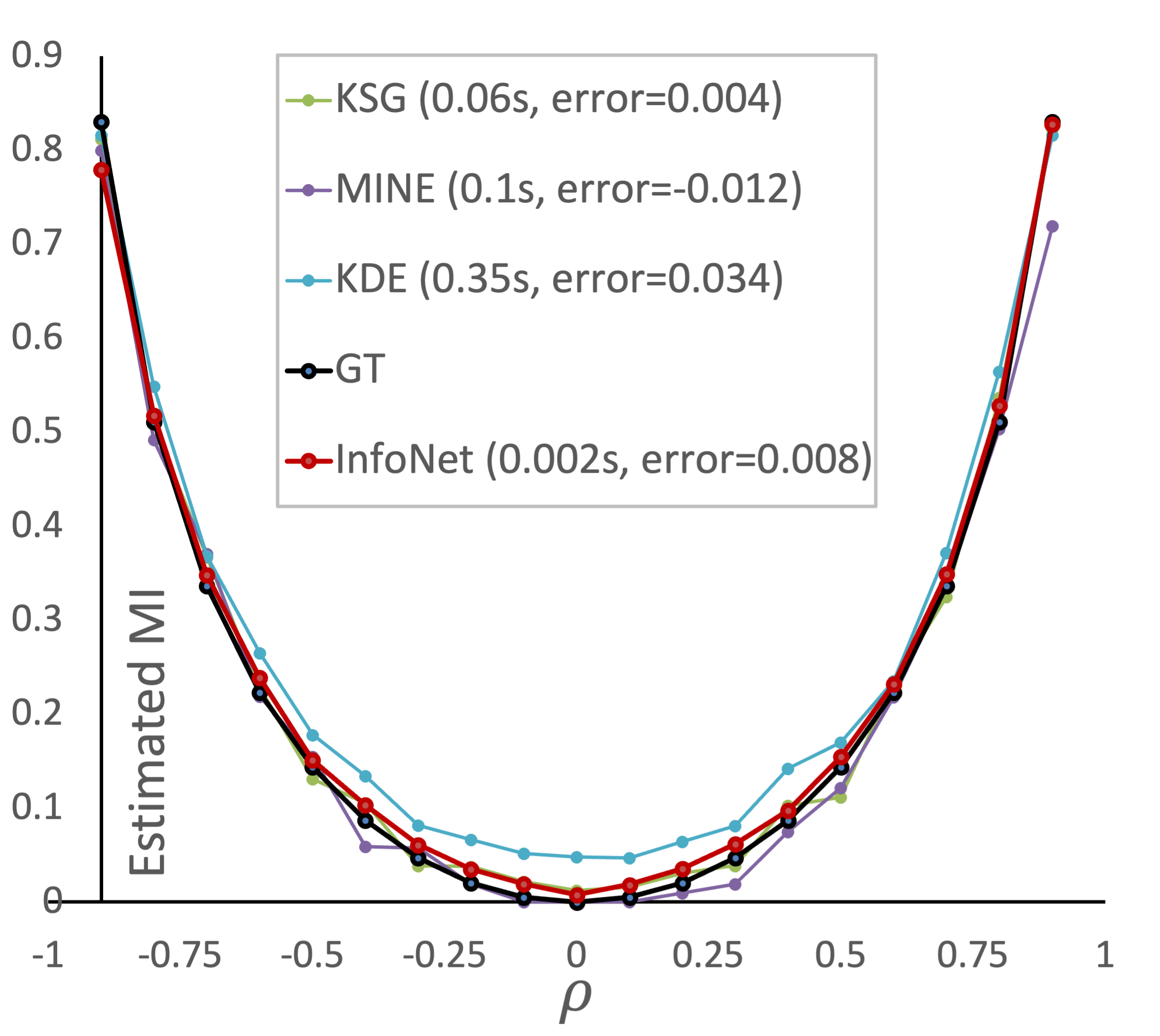}
    \vspace{-10pt}
    \caption{Comparison of MI estimates under Gaussian settings (runtime included).}
    \label{fig:single-gauss-test}
    \vspace{-18pt}
\end{figure}

We report the results and comparisons with three primary baselines. 
These baselines include: KSG \citep{kraskov2004estimating}, calculating MI by averaging k-nearest neighbor distances for entropy estimates; KDE \citep{silverman2018density}, which uses kernel functions to estimate joint and marginal densities, followed by MI computation through integration; and MINE \citep{MINE}, employing a similar dual formulation for MI estimation as InfoNet but resorts to optimizing a network for different distributions. All evaluations are conducted on an RTX 4090 GPU and an AMD Ryzen Threadripper PRO 5975WX 32-Core CPU.
\vspace{-2mm}

\begin{table*}[!t]
\setlength{\tabcolsep}{5pt}
\begin{footnotesize}
\caption{Error mean and variance of different MI estimators. Methods that do not rely on neural networks are highlighted in {\color{ballblue} Blue}, and those leveraging neural networks are colored {\color{asparagus} Green}. Numbers highlighted in bold represent the optimal performance achieved by neural estimators.}
\vspace{-3mm}
\label{tab:error_bar_results}
\begin{center}
\begin{tabular}{c | c c c c c c c c c c c}
\toprule
 & MI & 0.0 & 0.1 & 0.2 & 0.3 & 0.4 & 0.5 & 0.6 & 0.7 & 0.8 & 0.9 \\
\midrule

\multirow{5}{*}{\rotatebox{90}{Mean}} & {\color{ballblue} KSG}  & 0.001 & 0.001 & 0.004 & 0.006 & 0.008 & 0.009 & 0.012 & 0.015 & 0.016 & 0.014\\
& {\color{ballblue} KDE}  & 0.005 & 0.010 & -0.003 & -0.350 & -0.071 & -0.109 & -0.155 & -0.199 & -0.239 & -0.292  \\
& {\color{asparagus} MINE-500}  & \textbf{-0.003} & -0.058 & -0.116 & -0.173 & -0.228 & -0.294 & -0.344 & -0.399 & -0.431 & -0.485  \\
& {\color{asparagus} MINE-100}  & -0.008 & -0.092 & -0.173 & -0.251 & -0.336 & -0.420 & -0.504 & -0.584 & -0.658 & -0.742  \\
& {\color{asparagus}InfoNet}  & 0.010 & \textbf{0.004} & \textbf{0.008} & \textbf{-0.024} & \textbf{-0.040} & \textbf{-0.063} & \textbf{-0.082} & \textbf{-0.101} & \textbf{-0.124} &  \textbf{-0.138}    \\
\midrule

\multirow{5}{*}{\rotatebox{90}{Variance}} & {\color{ballblue}KSG}  & 2e-4 & 3e-4 & 4e-4 & 5e-4 & 6e-4 & 8e-4 & 9e-4 & 9e-4 & 1e-3 & 1e-3\\
& {\color{ballblue}KDE}  & 0.010 & 0.005 & 0.001 & 0.003 & 0.004 & 0.005 & 0.010 & 0.012 & 0.014 & 0.019  \\
& {\color{asparagus}MINE-500}  & 4e-5 & 0.001 & 0.004 & 0.008 & 0.013 & 0.018 & 0.027 & 0.039 & 0.052 & 0.060  \\
& {\color{asparagus}MINE-100}  & 4e-5 & 5e-4 & 0.002 & 0.005 & 0.009 & 0.012 & 0.017 & 0.025 & 0.033 & 0.040  \\
& {\color{asparagus}InfoNet}  & \textbf{1e-5} & \textbf{1e-4} & \textbf{3e-4} & \textbf{8e-4} & \textbf{0.001} & \textbf{0.002} & \textbf{0.004} & \textbf{0.005} & \textbf{0.007} &  \textbf{0.009}    \\
\bottomrule
\end{tabular}
\end{center}
\end{footnotesize}
\vspace{-3mm}
\end{table*}

\vspace{-1mm}
\paragraph{Test-Time Efficiency.}

We compare the time complexity of InfoNet with baseline methods on {\it new} distributions sampled with varying sequence lengths. 
The run times, averaged over 100 trials, are presented in Tab.~\ref{tab:time_results}, illustrating the efficiency of different approaches. 
For MINE, the parameters for test-time optimization are: a batch size of 100 and a learning rate of 0.001, while MINE-500 indicates 500 training iterations. 
For our approach, InfoNet-16 denotes the simultaneous estimation of 16 distributions in the batch mode. No training in InfoNet is needed.
The results demonstrate that InfoNet significantly outperforms other methods in processing speed for all sequence lengths tested, underscoring our method's efficiency in diverse settings.

\vspace{-3mm}
\paragraph{Sanity Check on Gaussian.}
We evaluate the MI estimation accuracy of different methods on Gaussian distributions. 
In this case, the MI of a (joint) distribution depends on their Pearson correlation coefficient $\rho$.
For a fair comparison, the MINE model is trained with a batch size of 500 for 500 steps at a learning rate of 0.001. 
The KSG method uses a neighborhood size of 5 for best performance. For each method, the number of data samples from the test distributions is 2000.
As shown in Fig.~\ref{fig:single-gauss-test}, InfoNet predicts MI values closer to the ground-truth MI compared to baseline methods, with mean error shown in the figure's legend.
We can see that InfoNet quantitatively achieves a similar error with KSG but is 30$\times$ faster.
When compared to MINE, InfoNet runs 50$\times$ faster, while achieving a 30\% improvement in accuracy.
This sanity check verifies that the proposed InfoNet has an optimal efficiency-accuracy tradeoff than others.
More results can be found in Appendix~\ref{sec:otherdis}.

\vspace{-3mm}
\paragraph{GMMs with Multiple Components.}

We also evaluate the above MI estimators 
on GMMs with multiple components, 
which is a more challenging but practical task.
Our test dataset is generated as follows: we define 10 MI levels, ranging from 0.0 to 0.9, and create random GMM distributions using the same data generation protocol. 
The MI of a sampled distribution is calculated using the Monte-Carlo Integration (MCI) method. 
A GMM distribution is saved to one of the 10 MI levels if its computed MI is within $\pm0.02$ of that level's value, while also recording the exact MI. 
The test data generation continues until each MI level has 1,000 test distributions. 
For each GMM distribution, 
we sample sequences with a length equal to 2000. These sequences' MI is then estimated using various methods. The mean error and variance for each method across different MI levels are summarized in Tab.~\ref{tab:error_bar_results}.

Accordingly, we can make the following observations:
1) Although traditional methods, e.g., KSG, perform relatively well in terms of mean error and variance, they cannot utilize neural networks for computational efficiency.
2) Among the neural methods (InfoNet and variants of MINE), our model achieves much smaller mean errors, and the prediction is more stable (in terms of variance) than MINE, which performs 100 and 500 gradient steps during the test-time training for different distributions.
The runtime for MINE-100 and MINE-500 are 0.17 and 0.991 seconds, respectively, while the runtime for InfoNet is 0.011 seconds.

\vspace{-4mm}
\paragraph{Mutual Correlation Order Accuracy.}

Now we report the results of various methods for the task of correlation order prediction with varying GMM components (e.g., $K$ ranges from 1 to 10), investigating how order accuracy changes with the difficulty of MI estimation.
As outlined in Sec.~\ref{para:setup}, we evaluate the estimated order of 2000 triplets per category ($K$) against the ground truth established by the MCI method. An order is deemed accurate if the estimated relationship ($\mathbb{I}(\rvx,\rvy)>\mathbb{I}(\rvx,\rvy')$ or $\mathbb{I}(\rvx,\rvy)\leq\mathbb{I}(\rvx,\rvy')$) aligns with the ground truth. The results are summarized in Tab.~\ref{tab:order_results}.

We can see that InfoNet consistently achieves higher order accuracy than the test-time optimization method (MINE), despite both utilizing neural networks. Furthermore, even as the difficulty of MI estimation increases ($K$ from 1 to 10), InfoNet reliably produces accurate order estimates between variables under different joint distributions. This underscores InfoNet's generalization as a neural estimator of correlation order for decision-making.
Performance curves comparing InfoNet and GT under different Gaussian components are shown in Appendix~\ref{sec:other-gmm-results} (Fig.~\ref{fig:sorted_result}).

\vspace{-3mm}

\begin{table*}[!t]
\begin{footnotesize}
\caption{Correlation order prediction accuracy 
of different MI estimators. 
Methods without neural networks are highlighted in {\color{ballblue} Blue}, and neural estimators are colored {\color{asparagus} Green}.
Performance is reported with various numbers of components in GMMs.}
\vspace{-3mm}
\label{tab:order_results}
\begin{center}
%\begin{small}
\begin{sc}
\begin{tabular}{c c c c c c c c c c c}
\toprule
 No. of Comps. & K=1 & K=2 & K=3 & K=4 & K=5 & K=6 & K=7 & K=8 & K=9 & K=10 \\
 
\midrule
{\color{ballblue}KSG}  & 98.7 & 99.0 & 98.2 & 98.0 & 97.9 & 97.7 & 97.6 & 97.5 & 97.0 & 97.3\\
{\color{ballblue}KDE}  & 97.4 & 97.7 & 97.9 & 97.5 & 97.9 & 97.8 & 97.0 & 97.4 & 97.4 & 97.4  \\
\midrule
{\color{asparagus}MINE-500}  & 98.5 & 91.2 & 90.8 & 87.2 & 84.5 & 83.7 & 81.2 & 79.6 & 81.3 & 78.1  \\
{\color{asparagus}MINE-100}  & 94.6 & 77.1 & 75.4 & 71.6 & 67.5 & 69.4 & 66.5 & 66.3 & 68.7 & 66.4\\
{\color{asparagus}MINE-10}  & 60.9 & 56.1 & 55.1 & 54.3 & 52.4 & 54.9 & 53.7 & 50.4 & 53.1 & 52.5 \\
{\color{asparagus}InfoNet}  & \textbf{99.8} & \textbf{99.5} & \textbf{99.0} & \textbf{99.2} & \textbf{99.1} & \textbf{99.2} & \textbf{99.0} & \textbf{99.2} & \textbf{99.3} &  \textbf{99.5}    \\

\bottomrule
\end{tabular}
\end{sc}
%\end{small}
\end{center}
\end{footnotesize}
\vspace{-5mm}
\end{table*}

\paragraph{High Dimensional Independence Testing.} \label{sec:high_dim_independent_test}

\begin{figure*}[ht]
    \centering
    \begin{subfigure}{.33\textwidth}
        \centering
        \includegraphics[width=\textwidth]{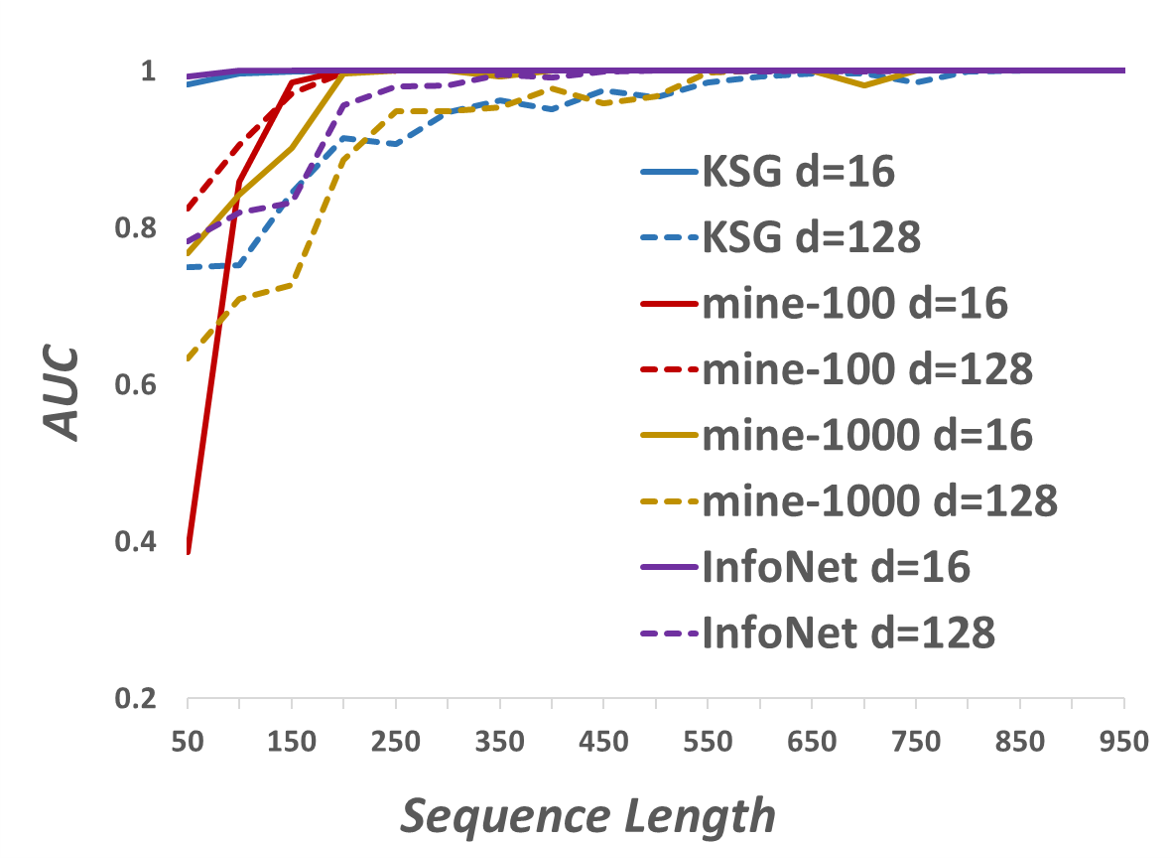}
        \caption{Correlation Type I}
        \vspace{-2mm}
    \end{subfigure}
    \begin{subfigure}{.33\textwidth}
        \centering
        \includegraphics[width=\textwidth]{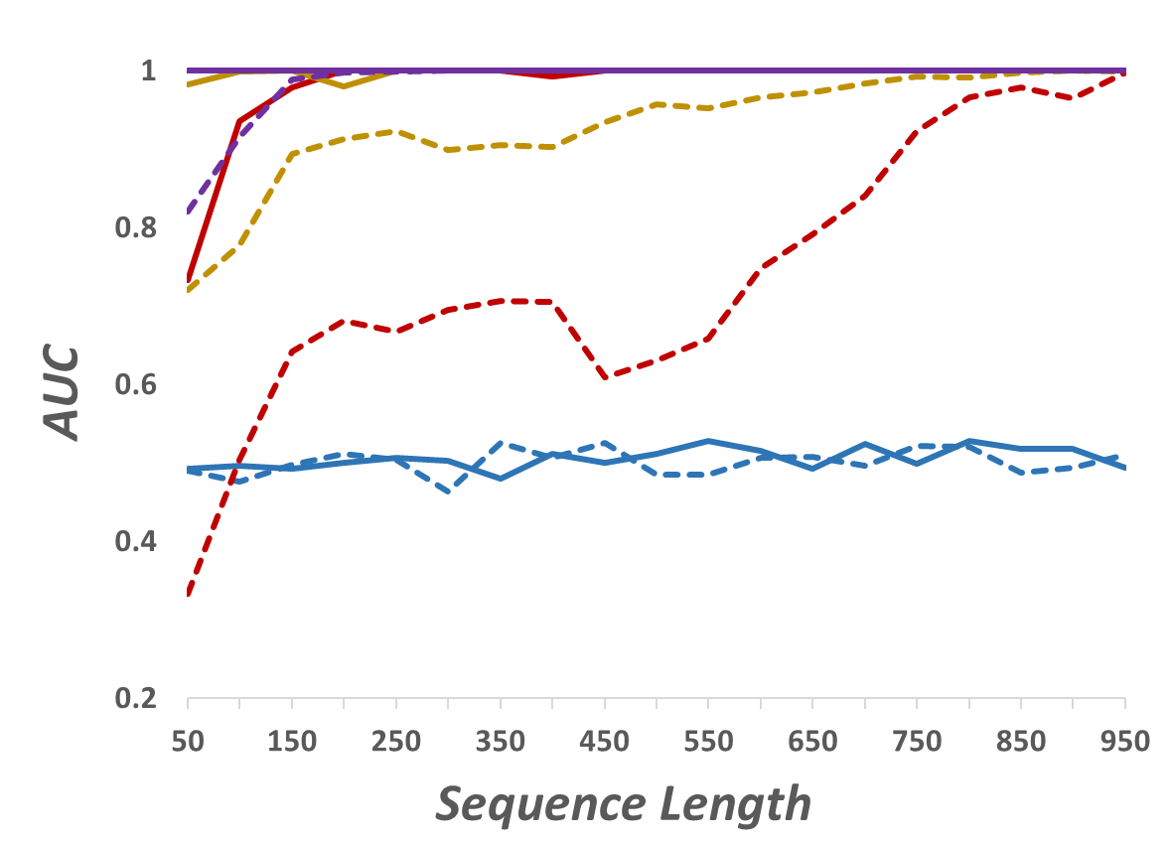}
        \caption{Correlation Type II}
        \vspace{-2mm}
    \end{subfigure}
    \begin{subfigure}{.33\textwidth}
        \centering
        \includegraphics[width=\textwidth]{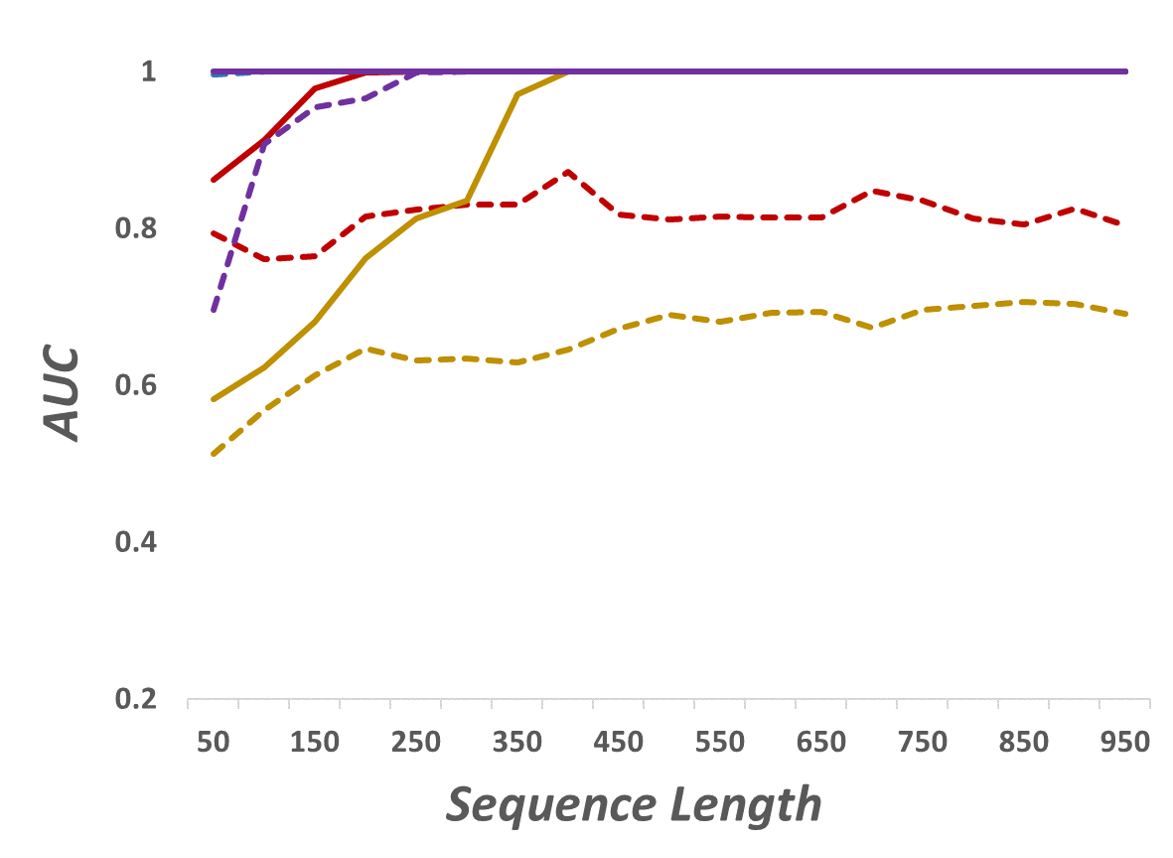}
        \caption{Correlation Type III}
        \vspace{-2mm}
    \end{subfigure}
    \vspace{-5mm}
    \caption{Independence testing under three types of data correlations. Each curve in the plots depicts the area under the curve (AUC) of the receiver operating characteristic (ROC) with respect to sequence length $n$. Four MI estimators are compared: InfoNet, KSG, MINE-100, and MINE-1000 (i.e., MINE trained with 100 and 1000 gradient steps during test-time optimization), each with two dimensions (16 and 128).
    The curves obtained by InfoNet (with the slicing technique) are constantly higher than the others, which demonstrates the effectiveness of InfoNet for dealing with high-dimensional data.}
    \label{fig:independent-test}
    \vspace{-10pt}
\end{figure*}

We assess InfoNet in dealing with high-dimensional random variables (distributions) by the independence testing proposed in Sec.~\ref{sec:eval-method}.
Even though trained with low-dimensional variables, we can easily adapt InfoNet for high-dimensional data by utilizing the slicing technique proposed in \citet{goldfeld2021sliced}.
Specifically, the sliced mutual information (SMI) $\mathbb{SdI}(X, Y)$ computed of high-dimensional $X$ and $Y$ with MI estimators (for low-dimensional variables) guarantees that $\mathbb{SdI}(X, Y) = 0$ implies $\mi(X, Y) = 0$, i.e., independence between $X$ and $Y$.

Fig.~\ref{fig:independent-test} shows the results of the proposed independence testing in three settings.
We report the area under the curve (AUC) of the receiver operating characteristic (ROC) for the slicing-empowered InfoNet, MINE, and KSG. 
The computation of SMI with InfoNet involves 1000 random projection steps in parallel. 
We obtain the high-dimensional test data with three types of data correlations (Appendix~\ref{sec:detail-express-independent}).
Each number on the curve is an average over ten trials. 
For each trial, 
the independence is evaluated on 100 pairs of RVs with different MI estimators, within which 50 are independent and the remaining 50 are dependent, balancing the labels. 
More details on how to generate high-dimensional joint distributions can be found in Appendix~\ref{sec:detail-express-independent}. 
To get the plot, we vary the sequence length $n$ and the RV dimension $d$ from 16 to 128, showing the variations of AUC of different methods. 
The plots in Fig.~\ref{fig:independent-test} demonstrate that InfoNet performs better than the other baselines in independence testing of high-dimensional random variables, verifying the effectiveness of InfoNet for dealing with high-dimensional data, especially with short sequences. 
More results on high-dimensional data can be found in Appendix~\ref{sec:other-highd-result}.

\vspace{-3mm}
\paragraph{Validation on Out-of-Domain Motion Data.} \label{sec:realworld-MCC}

We further evaluate our model's generalization to out-of-domain data.
Specifically,
we leverage InfoNet to perform mutual information estimation on motion trajectories of pixels in a video, and then use the estimated MI to perform object segmentation by thresholding the MIs.
If the estimation is correct, pixels
from the same object should be grouped together.

\begin{figure}[ht] 
   \centering 
   \includegraphics[width=0.82\linewidth]{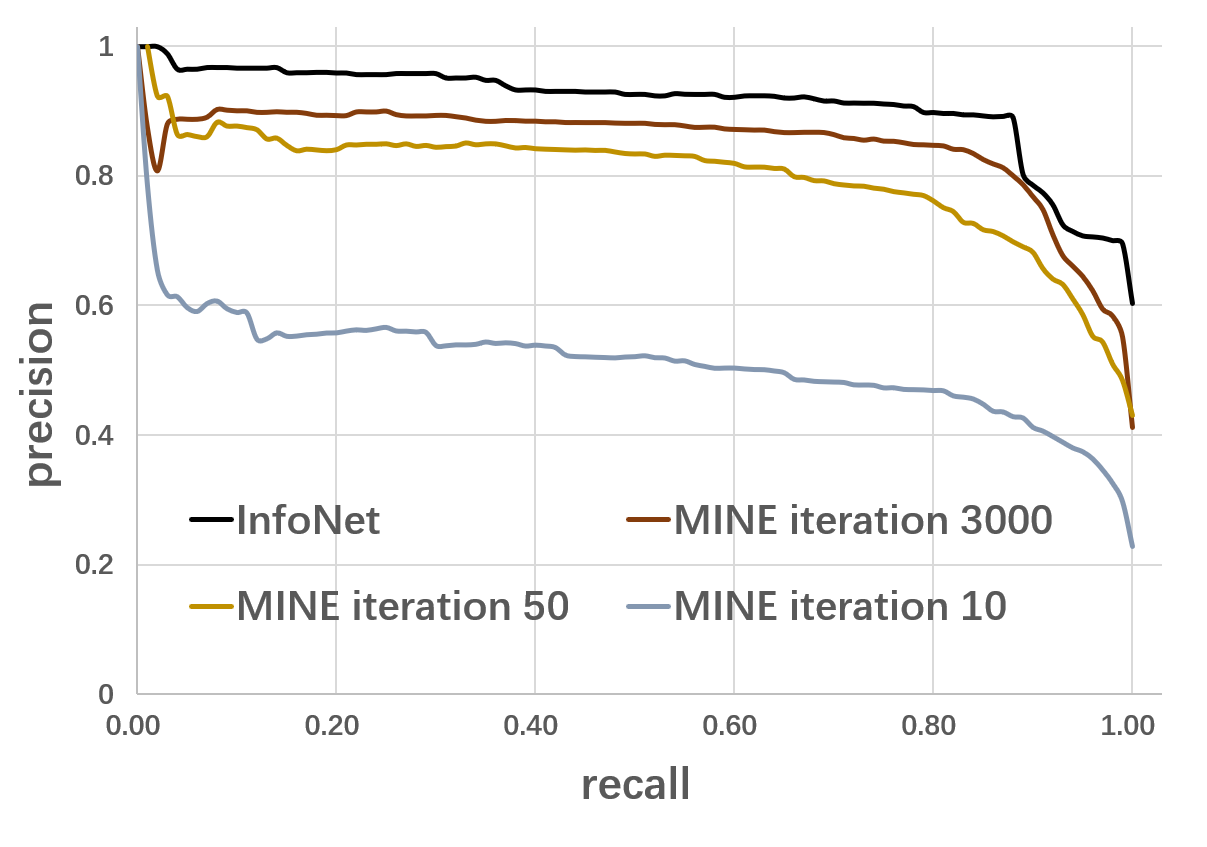}
   \vspace{-4mm}
   \caption{Precision-Recall curves of MI-based segmentation on Pointodyssey, verifying MI estimators' generalization to out-of-domain data.} 
   \label{fig:pr-graph} 
   \vspace{-6.6mm}
\end{figure}

We use the Pointodyssey dataset \citep{zheng2023pointodyssey}, consisting of long videos that provide rich ground-truth trajectories.
Given a pixel $P_{i}$ in the first frame of a video,
we can extract the locations of the corresponding pixels in all frames with the ground-truth trajectories. 
Then we utilize InfoNet to compute mutual information between the locations of any two pixels $P_{i}$ and $P_{i}$ in the first frame.
Since InfoNet does not resort to test-time optimization, the computation is efficient, e.g., less than 5 seconds for hundreds of pairs.
Instead of picking one threshold $\gamma$ for the grouping,
we vary it in an increment of $0.01$ to plot the precision-recall (P-R) curves with the help of the ground-truth object masks.
Fig.~\ref{fig:pr-graph} shows the P-R curves obtained for each MI estimation method on Pointodyssey. 
For more details, please refer to Appendix~\ref{sec:other-realworld-result}.
Results in Fig.~\ref{fig:pr-graph} demonstrate InfoNet's ability to achieve higher segmentation performance with sound generalization to out-of-domain data.

\vspace{-2mm}
\section{Related Works}\label{sec:related}

MI quantifies the statistical dependence between variables through a variety of nonparametric and parametric approaches. Nonparametric methods, such as K-Nearest Neighbors (KNN) and Kernel Density Estimation (KDE), estimate MI without assuming specific probability distributions \citep{reshef2011detecting,kinney2014equitability,khan2007relative,kwak2002input,kraskov2004estimating,pal2010estimation,gao2015efficient,gao2017estimating,runge2018conditional,lord2018geometric,moon1995estimation,steuer2002mutual,gretton2005kernel,kumar2021optimization}. These methods, however, suffer from limitations such as sensitivity to parameter choice, curse of dimensionality, computational complexity, and assumptions about continuity \citep{suzuki2008approximating,walters2009estimation,gao2018demystifying,mukherjee2020ccmi,fukumizu2007kernel,estevez2009normalized,bach2022information}. Binning methods and adaptive partitioning offer nonparametric alternatives but are constrained by bin/partition selection and the curse of dimensionality \citep{lugosi1996consistency,darbellay1999estimation,cellucci2005statistical,fernando2009selection,cakir2019hashing,marx2021estimating,thevenaz2000optimization,paninski2003estimation,knops2006normalized,tsimpiris2012nearest}. On the other hand, parametric methods assume specific distributions, such as Gaussian, but their accuracy is contingent upon correct assumptions and parameter estimation \citep{hulle2005edgeworth,gupta2010parametric,sugiyama2012density,gao2015estimating,ince2017statistical,suzuki2008approximating,walters2009estimation}.

Measuring and optimizing MI with limited sample sizes presents a challenge \citep{treves1995upward,mcallester2020formal}. Nevertheless, alternative measurements within a Reproducing Kernel Hilbert Space (RKHS) have demonstrated effectiveness in detecting statistical dependence \citep{gretton2005kernel}. Singular Value Decomposition (SVD) \citep{anantharam2013maximal,makur2015efficient}, Alternating Conditional Expectation (ACE) algorithm \cite{breiman1985estimating,buja1990remarks,huang2020sample,almaraz2020melanoma}, and rank correlation \citep{kendall1938new,klaassen1997efficient} are widely used conventional methods. Recently, neural network approaches have also been proposed \citep{xu2020maximal}.

Numerous works have addressed the scalable computation of MI and statistical dependences \cite{lopez2013randomized,mary2019fairness,goldfeld2021sliced,chen2022scalable}. Our proposed InfoNet offers an orthogonal alternative to these methods. Instead of striving for a more accurate approximation of the highly nonlinear MI or devising advanced yet computationally friendly correlation metrics, InfoNet focuses on MI estimation by encoding the optimization of its objectives into neural networks through pertaining, bypassing test-time optimization and conceptually allows for more efficient and accurate solutions to these complex correlation measures. The proposed method is also related to simulation-based intelligence \cite{cranmer2020frontier,ramon2021h3d}.
\vspace{-2mm}
\section{Discussion}\label{sec:discussion}

We present InfoNet, 
a novel neural network architecture 
for efficient MI estimation. 
Utilizing the attention mechanism 
and large-scale training, 
our approach circumvents time-consuming test-time optimization and demonstrates generalization capabilities. 
We extensively evaluated InfoNet's effectiveness on various distribution families and applications, emphasizing its efficiency-accuracy trade-off and order-preserving properties.
We validate InfoNet's potential in fields requiring real-time MI estimation and anticipate that our work will inspire further exploration of neuralizing the computation of MI and other information-theoretic quantities.
We also expect the proposed method and trained models can facilitate applications that require estimating a vast amount of correlation in a low time budget.

\section{Acknowledgment}

This work is supported by the HKU-100 Award and the HKU Seed Fund for Basic Research \# 2202100553.

% \newpage
% \input{sections/07_ethics}
% \input{sections/08_reproducibility}

\nocite{langley00}
% \paragraph{Impact Statements:} This paper presents work whose goal is to advance the field of Machine Learning. There are many potential societal consequences of our work, none of which we feel must be specifically highlighted here.
\bibliography{main}
\bibliographystyle{icml2024}

%%%%%%%%%%%%%%%%%%%%%%%%%%%%%%%%%%%%%%%%%%%%%%%%%%%%%%%%%%%%%%%%%%%%%%%%%%%%%%%
%%%%%%%%%%%%%%%%%%%%%%%%%%%%%%%%%%%%%%%%%%%%%%%%%%%%%%%%%%%%%%%%%%%%%%%%%%%%%%%
% APPENDIX
%%%%%%%%%%%%%%%%%%%%%%%%%%%%%%%%%%%%%%%%%%%%%%%%%%%%%%%%%%%%%%%%%%%%%%%%%%%%%%%
%%%%%%%%%%%%%%%%%%%%%%%%%%%%%%%%%%%%%%%%%%%%%%%%%%%%%%%%%%%%%%%%%%%%%%%%%%%%%%%
\newpage
\appendix
\onecolumn

\section{Appendix} \label{sec:appendix}

\subsection{Copulas} \label{sec:intro-copula}

To enhance the efficiency of MI estimation, we introduce the method called copula \citet{durante2010copula} during the data preprocessing stage. This approach is initiated based on a fundamental property of MI: given that $f, g: \mathbb{R} \rightarrow$ $\mathbb{R}$ are arbitrary strictly increasing functions, the following equation holds true:
\begin{equation}\label{eqn: mapping invariant}
\mathbb{I}\left(f(X), g(Y)\right)=\mathbb{I}\left(X, Y\right) .
\end{equation}

Specifically, drawing inspiration from \citep{pal2010estimation}, we choose mappings $f=F_X$ and $g=F_Y$, representing the cumulative distribution functions (CDFs) of random variables $X$ and $Y$. For continuous $F_X$ and $F_Y$, the marginal distribution uniformly spans $[0,1]$.

While the specific CDF of $X$ and $Y$ is not known in our situations, we employ the empirical CDF $\left(\widehat{F}_X, \widehat{F}_Y\right)$ as an alternative. Given a sequence $\mathbf{X}=\left({X}_1, {X}_2, \cdots, {X}_n\right)$ with length $n$, where each sample $X_i$, $i=1,\cdots,n$, originates from an unknown distribution, the empirical CDF is defined as follows:
\begin{equation}
    \widehat{F}_X(x)=\frac{1}{n}\operatorname{card}\left(\left\{i: 1 \leq i \leq n, x \leq X_i\right\}\right), \quad  x \in \mathbb{R},
\end{equation}
where $\operatorname{card}(\cdot)$ denotes the cardinality of the set. Note that while $\widehat{F}_X$ does not establish a bijection between $\mathbb{R}$ and the interval $[0,1]$, it is quite straightforward to create a bijection through interpolation while preserving the order of the sampling points. This ensures the invariance of MI, which remains unaffected by the transformation.

Our approach involves using the empirical CDF of $X$ and $Y$ to map them to a uniform distribution between $[0,1]$ prior to training and evaluation. 
In practice, this mapping process can be reduced to a simple sorting step: 
\begin{equation}
    {f(x) = \frac{1}{n} \operatorname{card}\left(\left\{i: 1 \leq i \leq n, x \leq X_i\right\}\right),\quad x=X_1,X_2,\cdots,X_n,}
\end{equation}
and
\begin{equation}
    g(y) = \frac{1}{n} \operatorname{card}\left(\left\{i: 1 \leq i \leq n, y \leq Y_i\right\}\right),\quad y=Y_1,Y_2,\cdots,Y_n,
\end{equation}
which are strictly increasing mappings that satisfy the requirement stated in \eqref{eqn: mapping invariant}.

\subsection{Sliced Mutual Information} \label{sec:intro-slicemi}

To estimate high-dimensional MI, we adopt the SMI concept \cite{goldfeld2021sliced}, which averages the MI between one-dimensional random projections of variables. Let $X$ and $Y$ be random variables with dimensions $d_x$ and $d_y$ respectively. SMI is thus the expected MI across these one-dimensional projections
\begin{equation}
\text{SMI}(X; Y) = \mathbb{E}_{\phi, \psi} \left[ \mathbb{I}(\phi(X); \psi(Y)) \right] = \frac{1}{S_{d_x-1} S_{d_y-1}} \oint_{\mathbb{S}^{d_x-1}} \oint_{\mathbb{S}^{d_y-1}} \mathbb{I}\left(\theta^{\top} X ; \phi^{\top} Y\right) \mathrm{d} \theta \mathrm{d} \phi
\end{equation}
Here, $\mathbb{S}^{d-1}$ denotes the $d$-dimensional sphere (whose surface area is designated by $S_{d-1}$), $\phi$ and $\psi$ are vectors used for linear projection from high-dimensional space to one-dimensional space, and $\mathbb{E}_{\phi, \psi}$ denotes the expectation over these projection functions.

While SMI typically yields lower values compared to MI, it retains many of the intrinsic properties of MI and exhibits a certain degree of correlation with it. This inter-connectedness is crucial, as it implies that while SMI offers a novel approach to handling high-dimensional data, it still adheres to the fundamental principles of MI, thereby ensuring consistency in its theoretical foundations and practical applications.

\subsection{Additional Results on GMMs with Multiple Components} \label{sec:other-gmm-results}

\begin{figure}[p] 
  \centering 
  \includegraphics[width=0.76\textwidth]{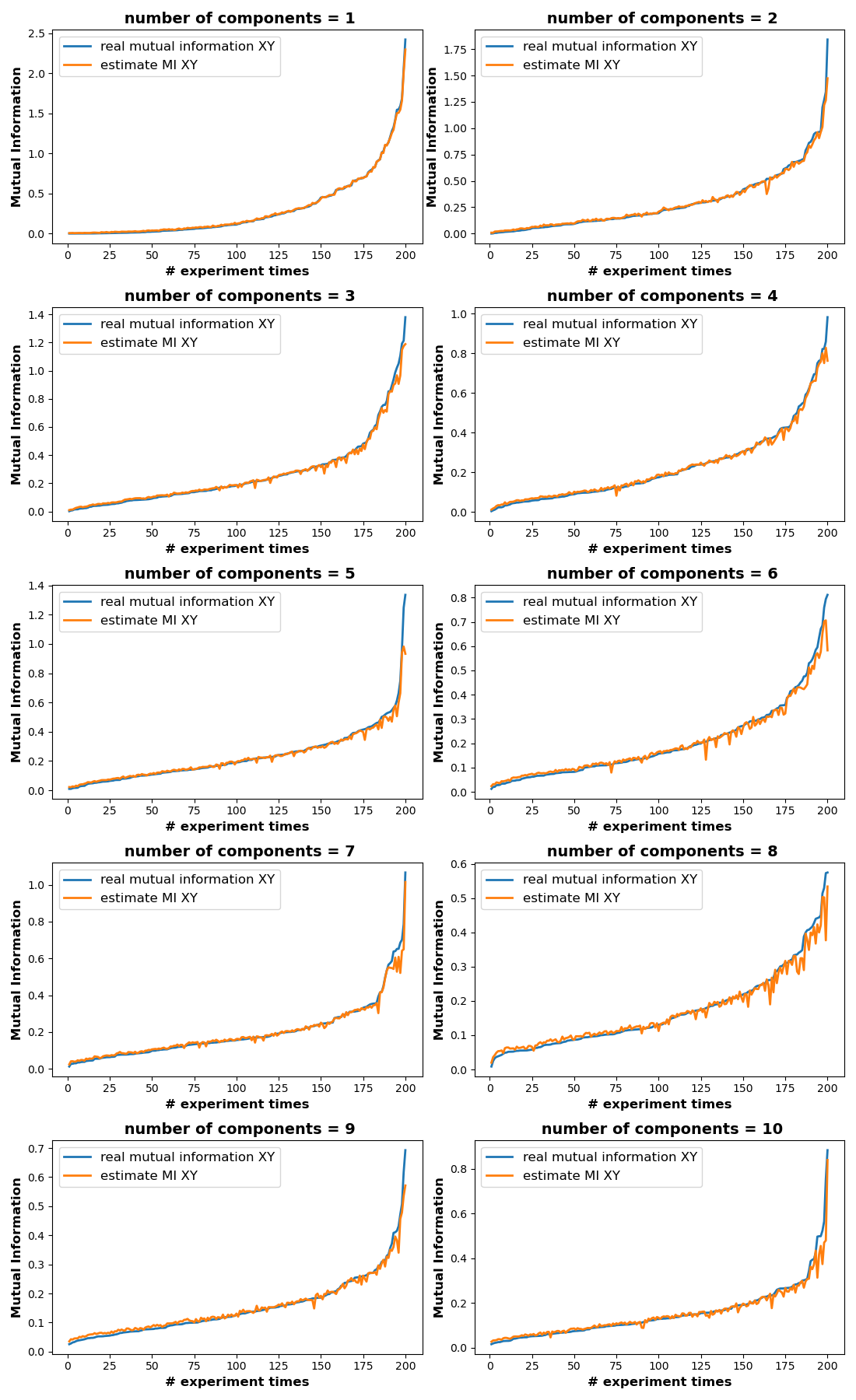} 
  \caption{Performance of our InfoNet model across various numbers of Gaussian components. The assessment is based on 200 randomly generated joint distributions in each category (number $K$ of GMM components), then sorted according to the value of ground-truth mutual information.} 
  \label{fig:sorted_result} 
\end{figure}

We demonstrate the capability of the InfoNet model in handling GMMs that comprise 1 to 10 Gaussian components. We evaluate the model's precision in estimating MI values by comparing these estimates with ground truth values. The results, depicted in Figure \ref{fig:sorted_result}, reveal that the InfoNet model accurately estimates MI on Gaussian mixture distributions, yielding estimates that are in close agreement with the ground truth values.

\subsection{Performance on Other Distributions} \label{sec:otherdis}

Paper \cite{czyz2023beyond} provides a diverse family of distributions with known ground truth
mutual information. We select three one-dimension distributions to test our InfoNet performance, note that our model has been only trained on GMM distributions and without any additional training.

\textbf{Half-Cube Map} Applying the half-cube homeomorphism $h(x)=|x|^{3 / 2} \operatorname{sign}(x)$ to Gaussian variables $X$ and $Y$, this could lengthen the tail. The transformation does not influence the ground truth value of MI.

\textbf{Asinh Mapping} Applying inverse hyperbolic sine function $\operatorname{asinh} x=$ $\log \left(x+\sqrt{1+x^2}\right)$ to shorten the tails, this transformation does not change the ground truth value of MI.

\textbf{Additive Noise} Let independent r.v. $X \sim \operatorname{Uniform}(0,1)$ and $N \sim \operatorname{Uniform}(-\varepsilon, \varepsilon)$, where $\varepsilon$ is the noise level. For $Y=X+N$, we could derive $\mathbf{I}(X ; Y)$ analytically.

\begin{figure}[h] 
  \centering 
  \includegraphics[width=0.92\textwidth]{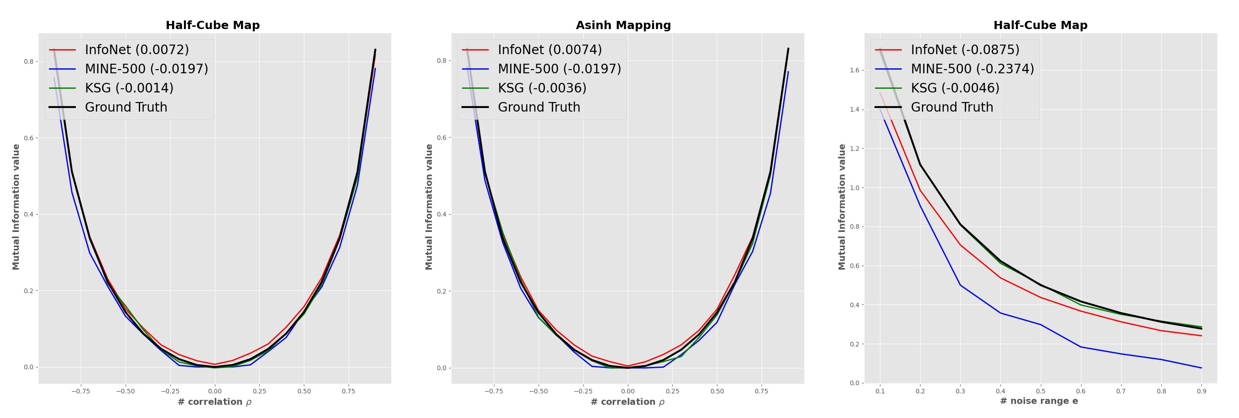} 
  \caption{Evaluation of performance on distribution other than GMM, comparing with MINE with 500 training iterations and KSG with nearest neighbor number $k=1$.} 
  \label{fig:other_dis} 
  \vspace{-3mm}
\end{figure}

Fig.~\ref{fig:other_dis} shows our result on other distributions despite the Mixture of Gaussian distributions. Due to the introduction of the copula, our model can suit different monotonic transformations well and produce good estimations for Half-Cube Map and Asinh Mapping. Also, our model performs well on Additive noise, evidencing good generalization ability as we do not train it on any uniform distributions and additive noise.

\subsection{Three types of dependencies between $X$ and $Y$}
\label{sec:detail-express-independent}

Below are three different relationships between $X$ and $Y$ in high dimensional independence test in sec.~\ref{sec:high_dim_independent_test}.

(a) \textbf{One feature (linear)}: $X, Z \sim \mathcal{N}\left(0, \mathrm{I}_d\right)$ i.i.d. and $Y=\frac{1}{\sqrt{2}}\left(\frac{1}{\sqrt{d}}\left(\mathbf{1}^{\top} X\right) \mathbf{1}+Z\right)$, where $\mathbf{1}:=$ $(1, \ldots ,1)^{\top} \in \mathbb{R}^d$.

(b) \textbf{Two features}: $X, Z \sim \mathcal{N}\left(0, \mathrm{I}_d\right)$ i.i.d. and $Y_i=\frac{1}{\sqrt{2}} \begin{cases}\frac{1}{d}\left(\mathbf{1}_{\lfloor d / 2\rfloor} 0 \ldots 0\right)^{\top} X+Z_i, & i \leq \frac{d}{2} \\ \frac{1}{d}\left(0 \ldots 0 \mathbf{1}_{\lceil d / 2\rceil}\right)^{\top} X+Z_i, & i>\frac{d}{2} \text {. }\end{cases}$

(c) \textbf{Independent coordinates}: $X, Z \sim \mathcal{N}\left(0, \mathrm{I}_d\right)$ i.i.d. and $Y=\frac{1}{\sqrt{2}}(X+Z)$.

\subsection{Additional Result On High Dimension} \label{sec:other-highd-result}

Similarly, we validate the capability of InfoNet in classifying the correct correlation order on $d$-dimensional Gauss distributions: $(X,Y) = \left( (X^1, X^2, \dotsm X^d), (Y^1, Y^2, \dotsm Y^d) \right)  \sim \mathcal{N}(\boldsymbol{\mu}, \Sigma)$.

This result shows that our InfoNet model reaches high accuracy and still costs low time complexity. Since our model allows parallel computing on multiple GPUs, it can compute the MI of multiple projected variables in one feed-forward process. 

\begin{table*}[!h]
\caption{Correlation order accuracy of different MI estimators. Methods that do not rely on neural networks are highlighted in {\color{ballblue} Blue}, and those leveraging neural networks are colored {\color{asparagus} Green}.
MINE-100 means training MINE method for 100 iterations, InfoNet-100 means we do 100 times random projection to get an average.}
\vspace{-3mm}
\label{tab:high_dim_order_results}
\begin{center}
\begin{sc}
\begin{tabular}{c c c c c c c c c c}
\toprule
 Dimensions & 2 & 3 & 4 & 5 & 6 & 7 & 8 & 9 & 10 \\
 
\midrule
{\color{ballblue}KSG}  & 94.4 & 95.5 & 91.8 & 92 & 94.1 & 93.6 & 94.1 & 94.1 & 94.2 \\
{\color{ballblue}Energy Distance}  & 49.6 & 51.2 & 52.2 & 51.5 & 52.5 & 49.6 & 48.7 & 50.2 & 51.3 \\
\midrule
{\color{asparagus}MINE-100}  & 78.5 & 82.1 & 86.7 & 84.7 & 88.4 & 89.8 & 90.1 & 90.4 & 90 \\
{\color{asparagus}MINE-1000} & 93.6 & 93.9 & 94.4 & 94.3 & 91.6 & 91.7 & 89.5 & 91 & 90.3   \\
{\color{asparagus}MINE-5000}  & 96.2 & \textbf{97} & \textbf{97} & 96.2 & 94.9 & 94.2 & 93.2 & 92.8 & 93   \\
{\color{asparagus}InfoNet-100}  & 93.7 & 94.6 & 94.4 & 95.7 & 93.3 & 95.8 & 95.8 & 95.4 & 93.8\\
{\color{asparagus}InfoNet-500}  & 94.9 & 93.7 & 95.7 & 95.8 & 97.1 & 96.4 & 97.2 & 97.8 & 96.8 \\
{\color{asparagus}InfoNet-1500}  & \textbf{97.7} & 96.4 & 96.2 & \textbf{97.9} & \textbf{97.4} &  \textbf{98.1} & \textbf{98.2} & \textbf{97.3} &  \textbf{98.3} \\

\bottomrule
\end{tabular}
\end{sc}
\end{center}
\vspace{-5mm}
\end{table*}

\subsection{Results of Validation on Out-of-Domain Motion Data}\label{sec:other-realworld-result}

In this section, we provide detailed results of the experiments on motion data.

It is worth noting that certain trajectories provided may contain unreasonable values such as "inf" or "-50000". To address this issue in the dataset, we apply a filtering process to ensure that only points appearing throughout the entire video are considered for analysis.

\begin{figure*}[h]
    \centering
    \begin{subfigure}{0.45\textwidth}
        \includegraphics[width=\linewidth]{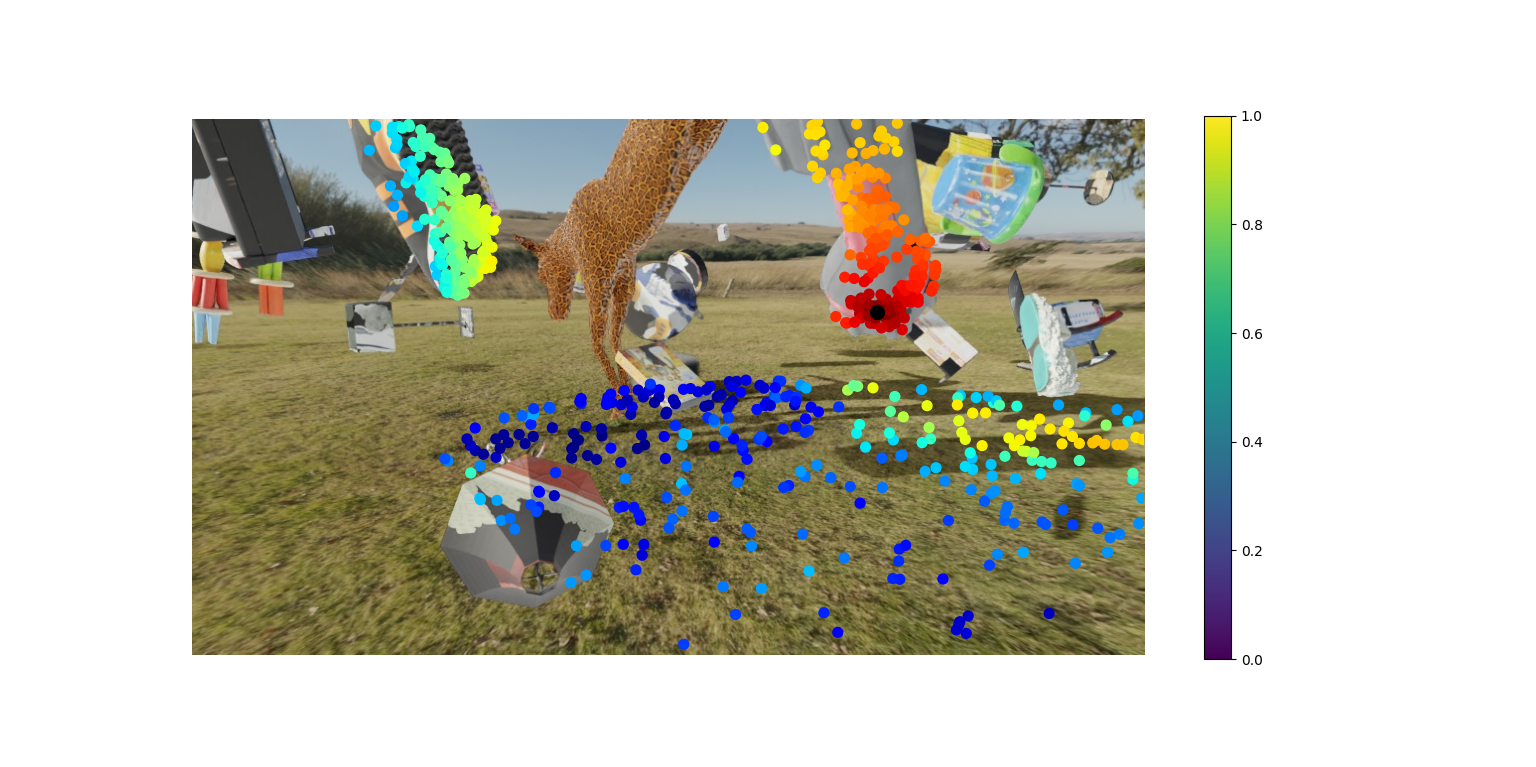}
        \vspace{-8mm}
        \caption{Estimated Mutual Information with point in object 1 (highlighted black).}
        \label{fig:video_mi1}
    \end{subfigure}
    \hspace{5mm}
    \begin{subfigure}{0.45\textwidth}
        \includegraphics[width=\linewidth]{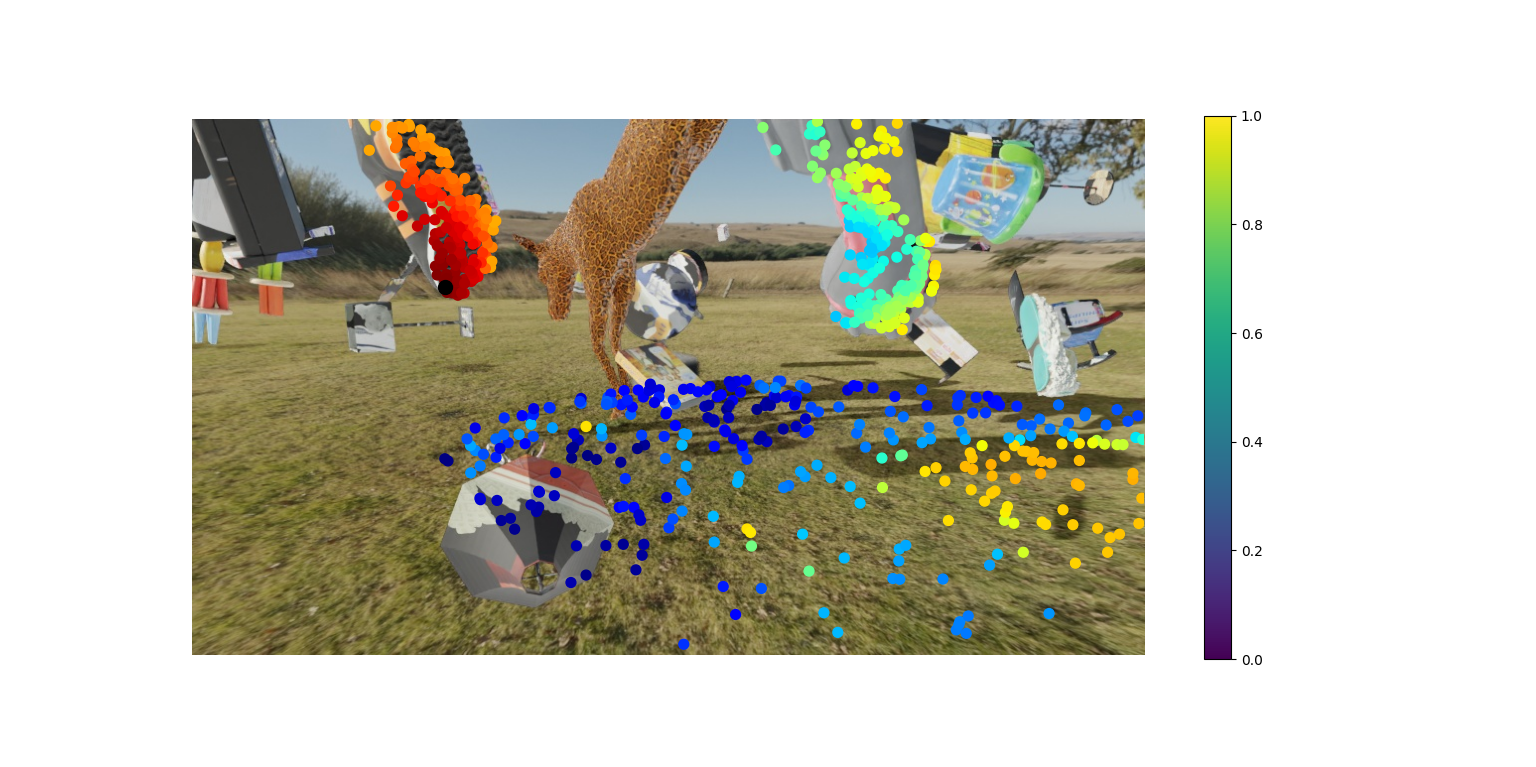}
        \vspace{-8mm}
        \caption{Estimated Mutual Information with point in object 2 (highlighted black).}
        \label{fig:video_mi2}
    \end{subfigure}
    \caption{Visualization results using InfoNet model.}
    \label{fig:motion-2point}
    \vspace{-3mm}
\end{figure*}

\begin{figure}[h] 
  \centering 
  \includegraphics[width=0.95\textwidth]{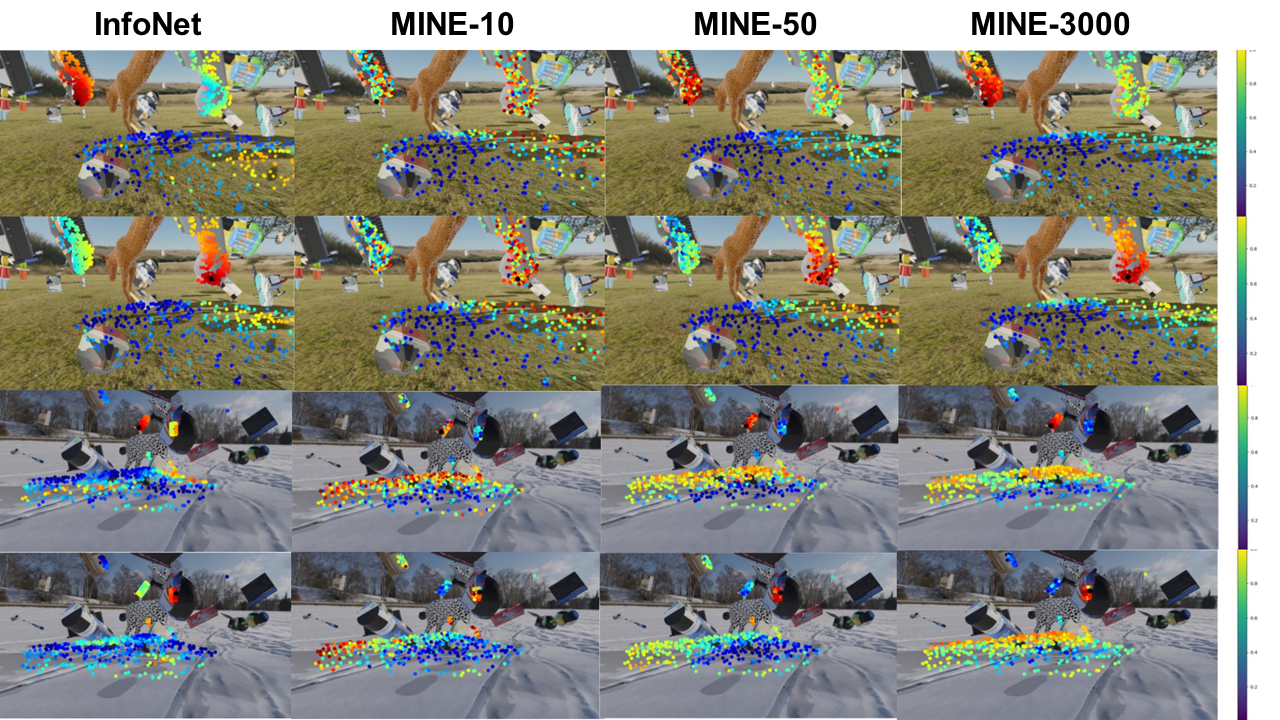} 
  \includegraphics[width=0.95\textwidth]{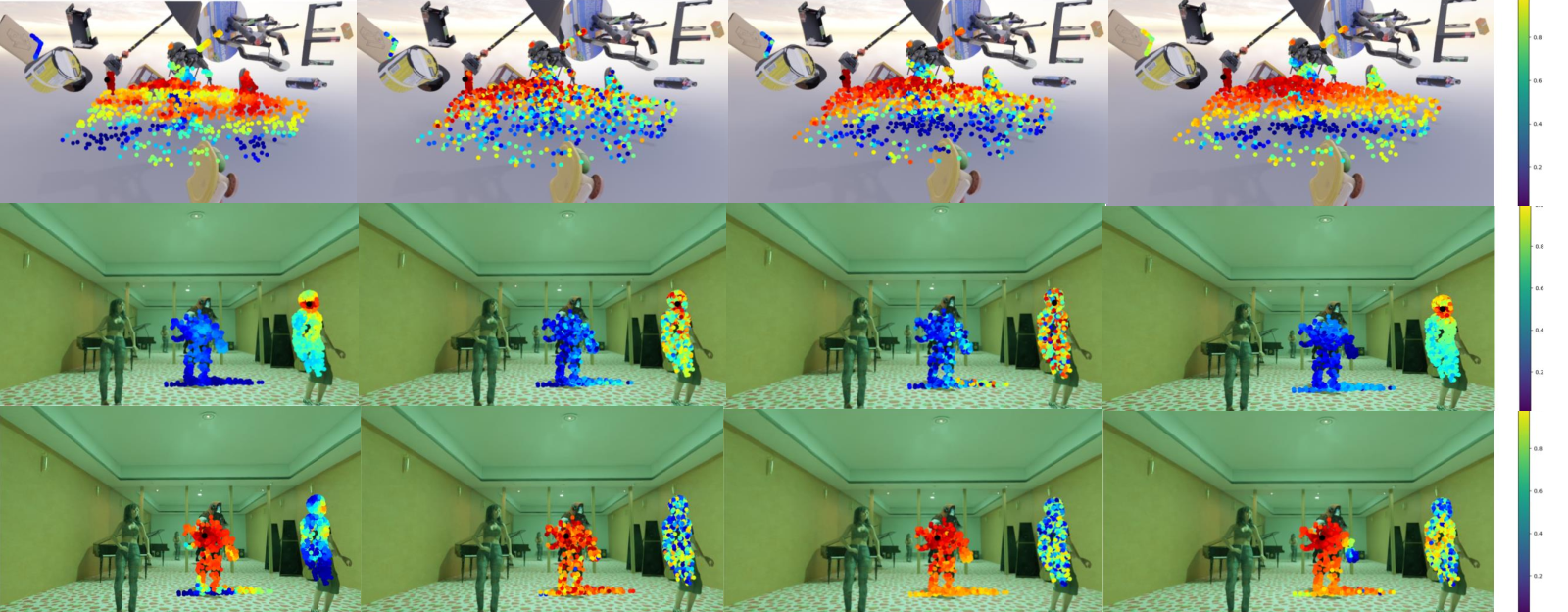} 
  \caption{Visual comparison of estimated MI of pixel locations between our model and MINE on the video datasets. Large value in red while small value in blue. From left to right: InfoNet (no test-time optimization), MINE (test-time gradient steps 10), MINE (test-time gradient steps 50), and MINE (test-time gradient steps 3000).} 
  \label{fig:video_visible} 
\end{figure}

\begin{figure}[h]
    \centering
    \begin{subfigure}[b]{0.48\textwidth}
            \includegraphics[width=\textwidth]{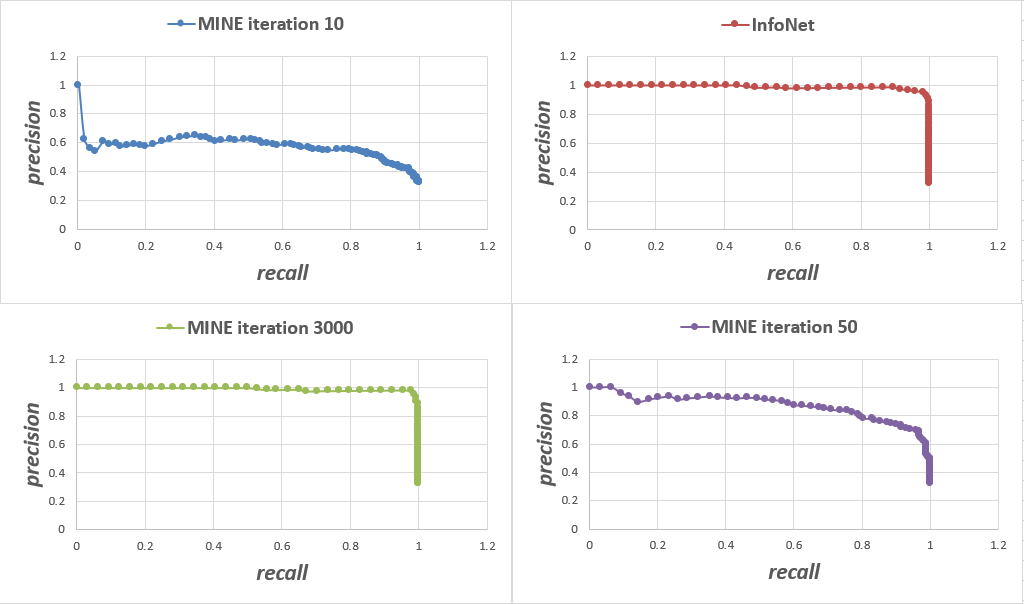}
            \caption{}
            \label{fig:individual precision recall curve 1}
        \end{subfigure}
        \hfill % or \hspace{5em}
    \begin{subfigure}[b]{0.48\textwidth}
            \includegraphics[width=\textwidth]{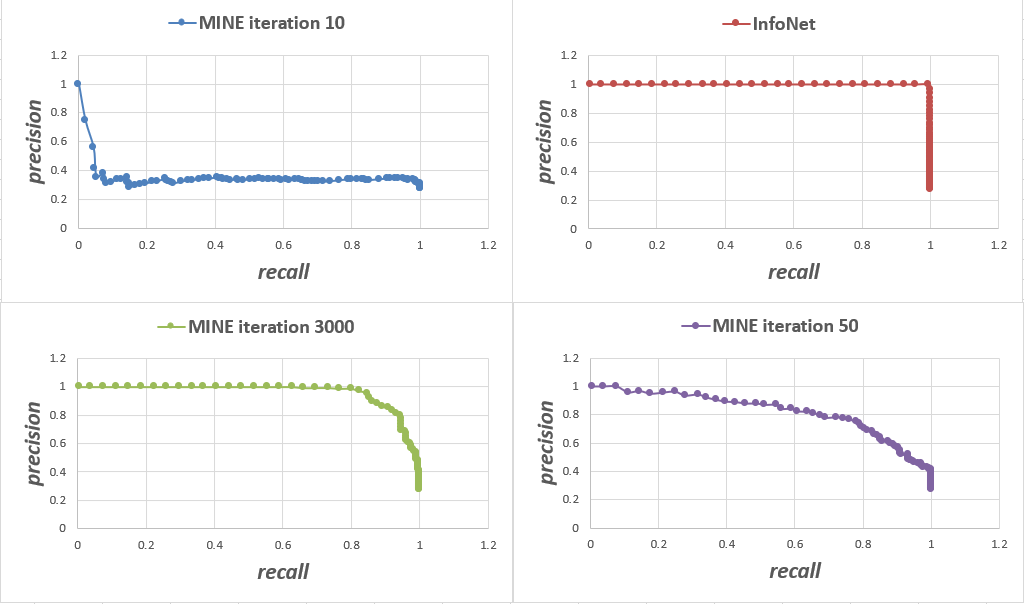}
            \caption{}
            \label{fig:individual precision recall curve 2}
        \end{subfigure}
        \hfill % or \hspace{5em}
    \begin{subfigure}[b]{0.48\textwidth}
            \includegraphics[width=\textwidth]{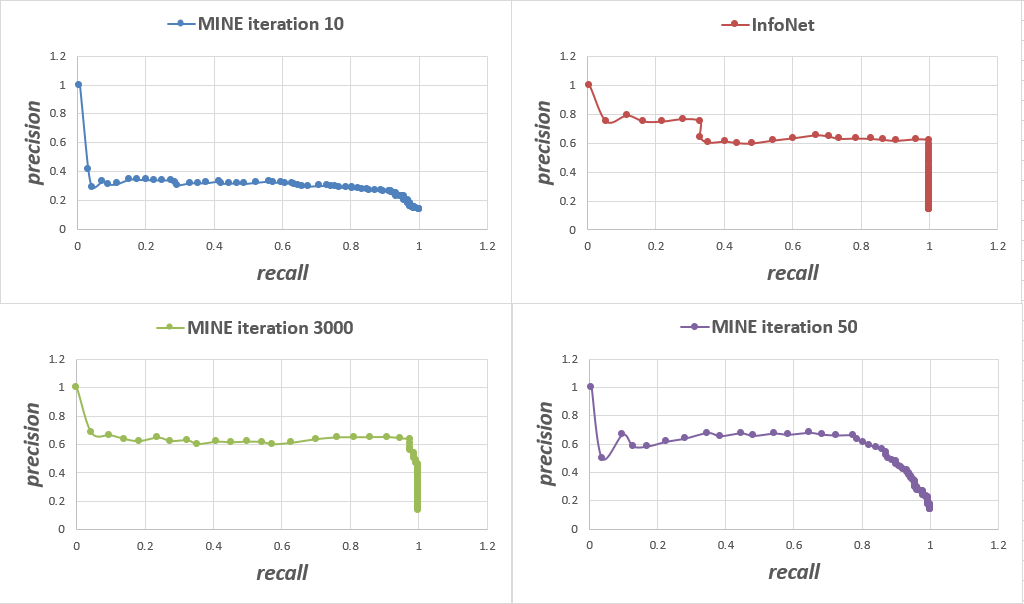}
            \caption{}
            \label{fig:individual precision recall curve 3}
        \end{subfigure}
        \hfill % or \hspace{5em}
    \begin{subfigure}[b]{0.48\textwidth}
            \includegraphics[width=\textwidth]{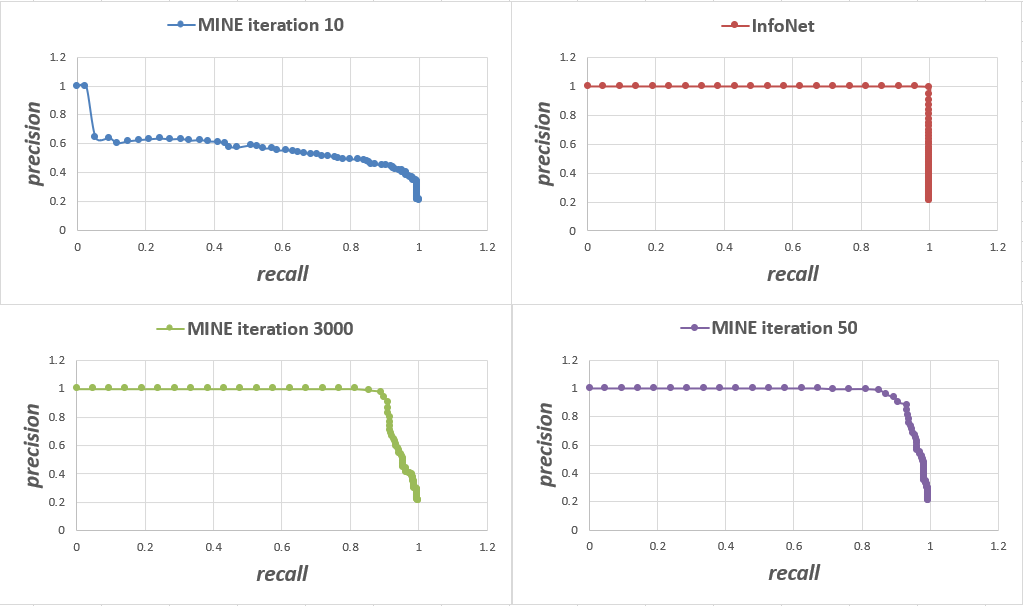}
            \caption{}
            \label{fig:individual precision recall curve 4}
        \end{subfigure}
        \hfill % or \hspace{5em}    
    \begin{subfigure}[b]{0.48\textwidth}
            \includegraphics[width=\textwidth]{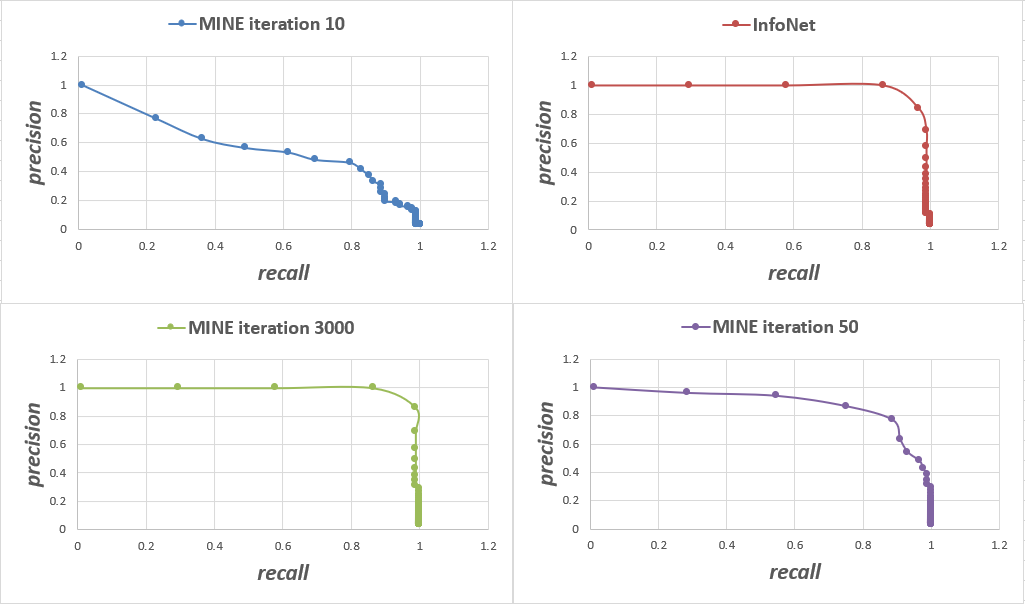}
            \caption{}
            \label{fig:individual precision recall curve 5}
        \end{subfigure}
    \begin{subfigure}[b]{0.48\textwidth}
            \includegraphics[width=\textwidth]{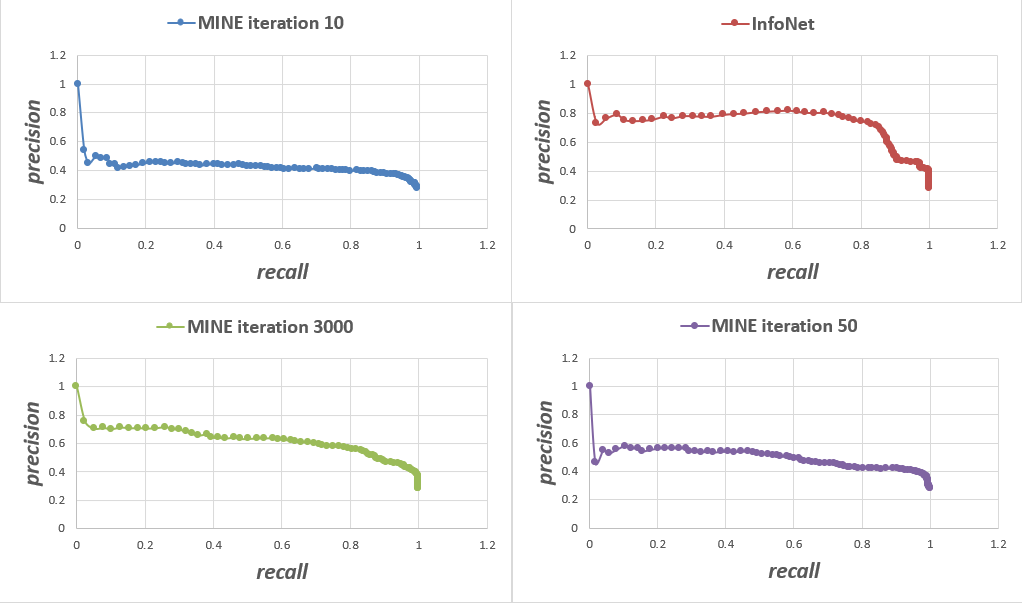}
            \caption{}
            \label{fig:individual precision recall curve 6}
        \end{subfigure}
        \hfill % or \hspace{5em} 
    \begin{subfigure}[b]{0.48\textwidth}
            \includegraphics[width=\textwidth]{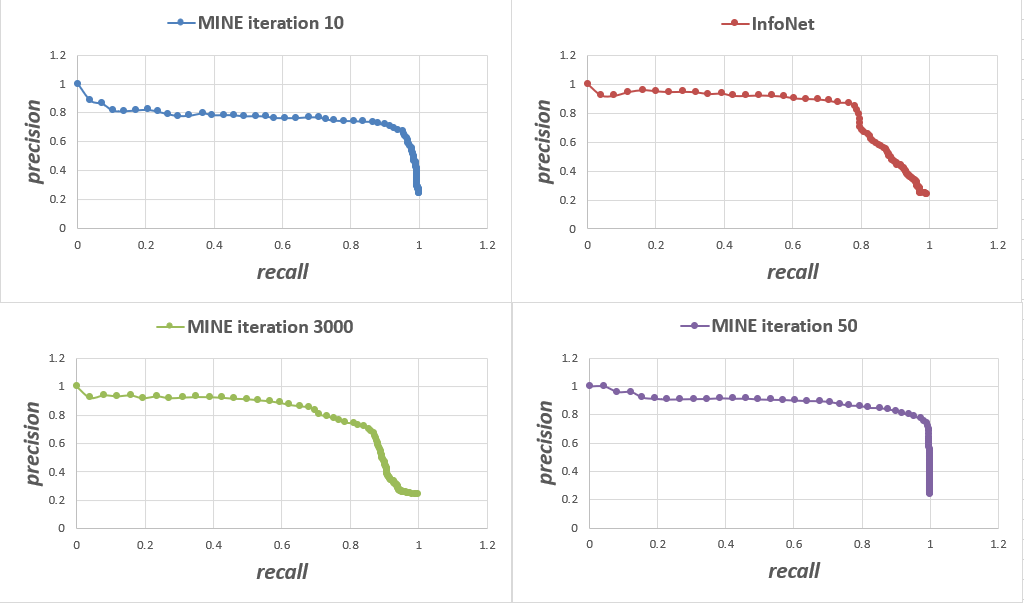}
            \caption{}
            \label{fig:individual precision recall curve 7}
        \end{subfigure}
        \hfill % or \hspace{5em}     
  
    \caption{Individual PR graph of our model and MINE. In the experiments conducted on video datasets, InfoNet exhibited notably high stability compared to MINE.}
    \label{fig:pr-graphs}
\end{figure}

\begin{figure}[h]
    \centering
    \begin{subfigure}[b]{0.48\textwidth}
            \includegraphics[width=\textwidth]{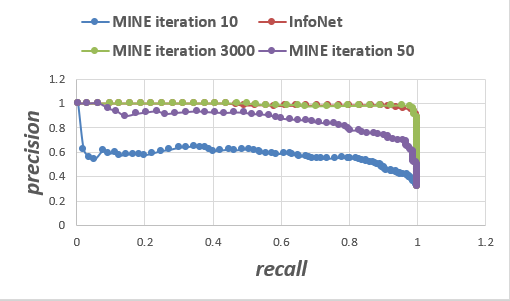}
            \caption{}
            \label{fig:Joint precision recall curve 1}
        \end{subfigure}
        \hfill % or \hspace{5em}
    \begin{subfigure}[b]{0.48\textwidth}
            \includegraphics[width=\textwidth]{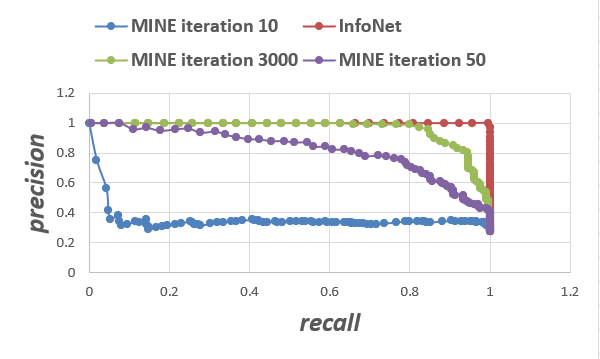}
            \caption{}
            \label{fig:Joint precision recall curve  2}
        \end{subfigure}
        \hfill % or \hspace{5em}
    \begin{subfigure}[b]{0.48\textwidth}
            \includegraphics[width=\textwidth]{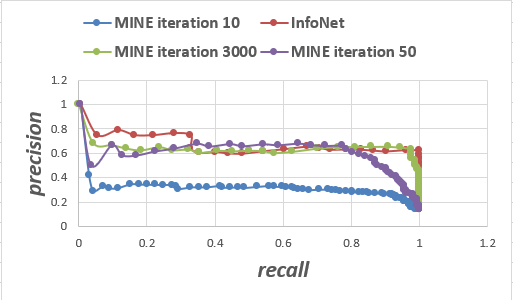}
            \caption{}
            \label{fig:fig:Joint precision recall curve  3}
        \end{subfigure}
        \hfill % or \hspace{5em}
    \begin{subfigure}[b]{0.48\textwidth}
            \includegraphics[width=\textwidth]{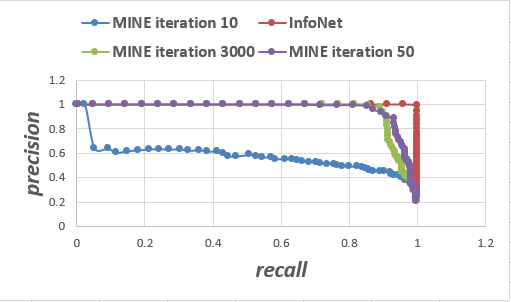}
            \caption{}
            \label{fig:fig:Joint precision recall curve  4}
        \end{subfigure}
        \hfill % or \hspace{5em}    
    \begin{subfigure}[b]{0.48\textwidth}
            \includegraphics[width=\textwidth]{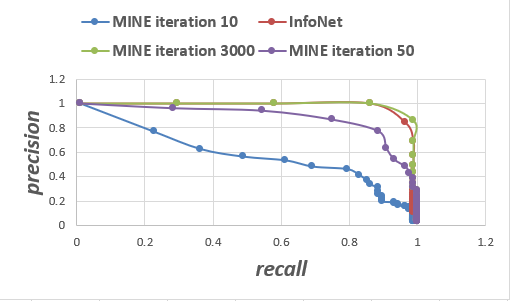}
            \caption{}
            \label{fig:fig:Joint precision recall curve  5}
        \end{subfigure}
    \begin{subfigure}[b]{0.48\textwidth}
            \includegraphics[width=\textwidth]{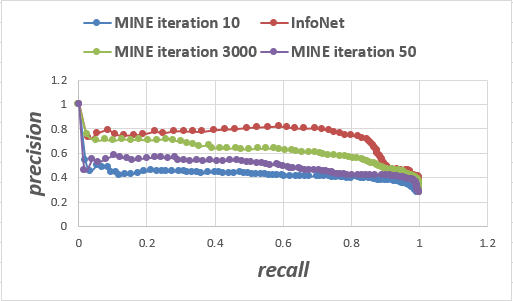}
            \caption{}
            \label{fig:fig:Joint precision recall curve  6}
        \end{subfigure}
        \hfill % or \hspace{5em} 
    \begin{subfigure}[b]{0.48\textwidth}
            \includegraphics[width=\textwidth]{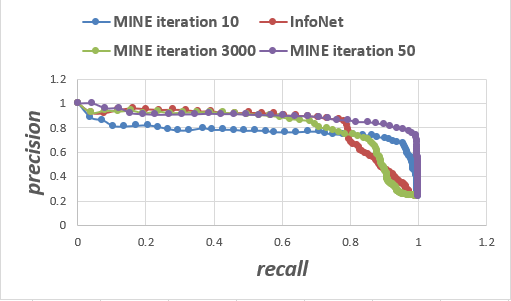}
            \caption{}
            \label{fig:fig:Joint precision recall curve  7}
        \end{subfigure}
        \hfill % or \hspace{5em}     
  
    \caption{Comparison between PR graphs of our model and MINE. In the same video dataset, InfoNet consistently exhibits superior performance compared to MINE.}
    \label{Joint precision recall curves}
    
\end{figure}

Fig.~\ref{fig:motion-2point} and Fig.~\ref{fig:video_visible} show the visualization of estimated mutual information between one selected point and other points in the videos. Fig.~\ref{fig:pr-graphs}
presents the individual PR curves for each object, while Fig \ref{Joint precision recall curves} provides the comparison of PR curves across different methods on each object.

\vspace{-3mm}

\subsection{Smoothing the Lookup Table}

We have tried two smoothing techniques to enhance the smoothness of the lookup table, and choosing method 1 in our final presented results.

\begin{figure}[h] 
  \centering 
  \includegraphics[width=0.8\textwidth]{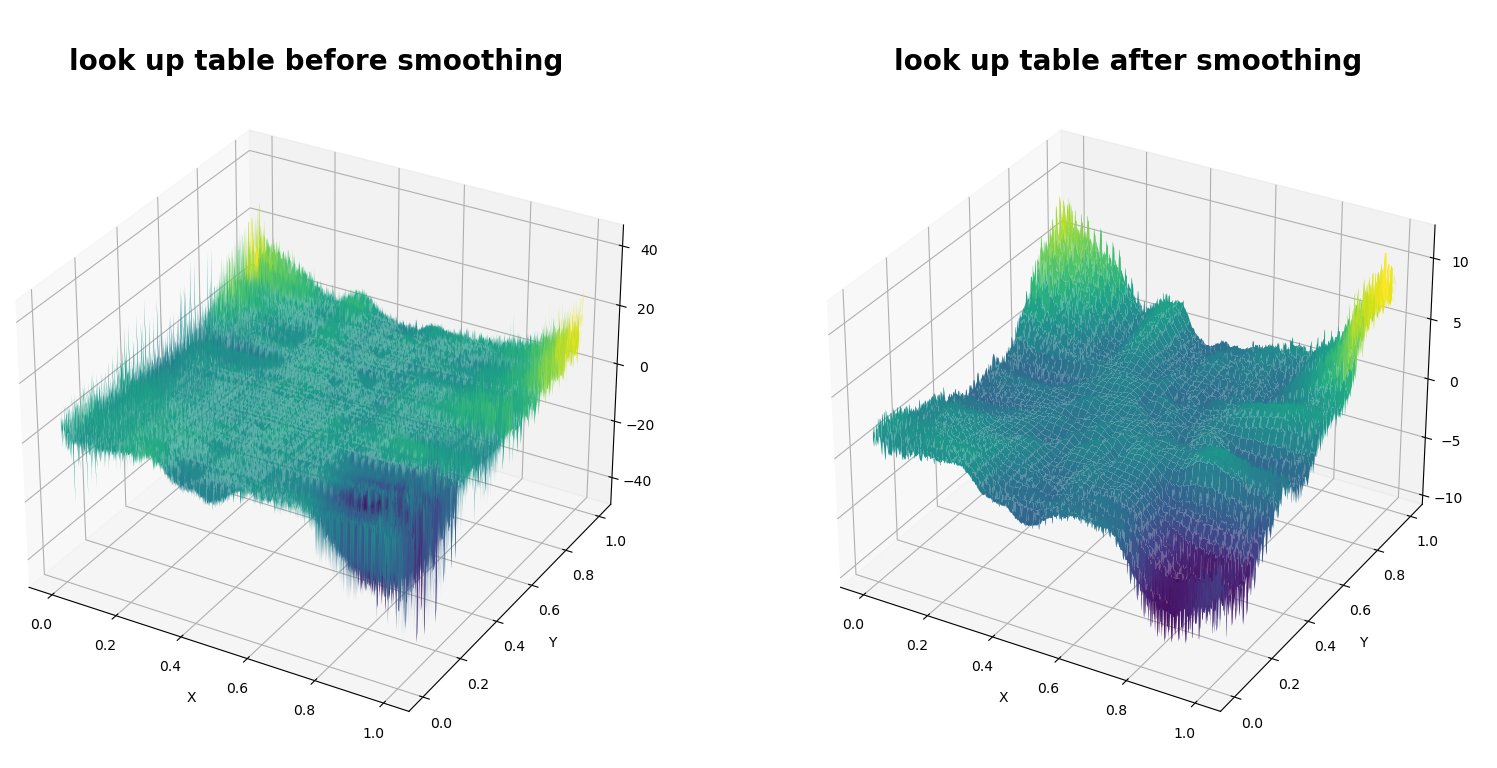} 
  \caption{Comparison between the un-smoothed lookup table and smoothed lookup table.} 
  \label{fig:smooth_table} 
\end{figure}

Method 1: We apply a convolution layer with a non-learnable Gaussian kernel with size 15 and sigma 3 on the lookup-table layer. Fig.~\ref{fig:smooth_table} shows the visualization results of adding the Gaussian smooth kernel. The table is indeed much smoother than not applying the kernal.

Method 2: We add a penalty term to punish the jumps in values between adjacent points using the Laplacian operator. We add $\mathcal{L}_{smooth} = \alpha \  |\text{Laplacian}( \text{lookup-table} )|$ after Eq. \ref{eq:explict-mi-loss}, where $\text{Laplacian}( \text{lookup-table} )$ can be obtained by applying a convolution layer on the lookup-table using Laplacian kernel: 
\begin{equation}
\text{Laplacian Kernel} =
\begin{bmatrix}
0 & 1 & 0 \\
1 & -4 & 1 \\
0 & 1 & 0
\end{bmatrix}
\end{equation}

\subsection{Data Distributions} \label{sec:data-distribution}

In this section, we provide several plots to visualize the sequences sampled from randomly generated Gaussian mixture distributions used for training.

\begin{figure}[!h] 
  \centering 
  \includegraphics[width=0.8\textwidth]{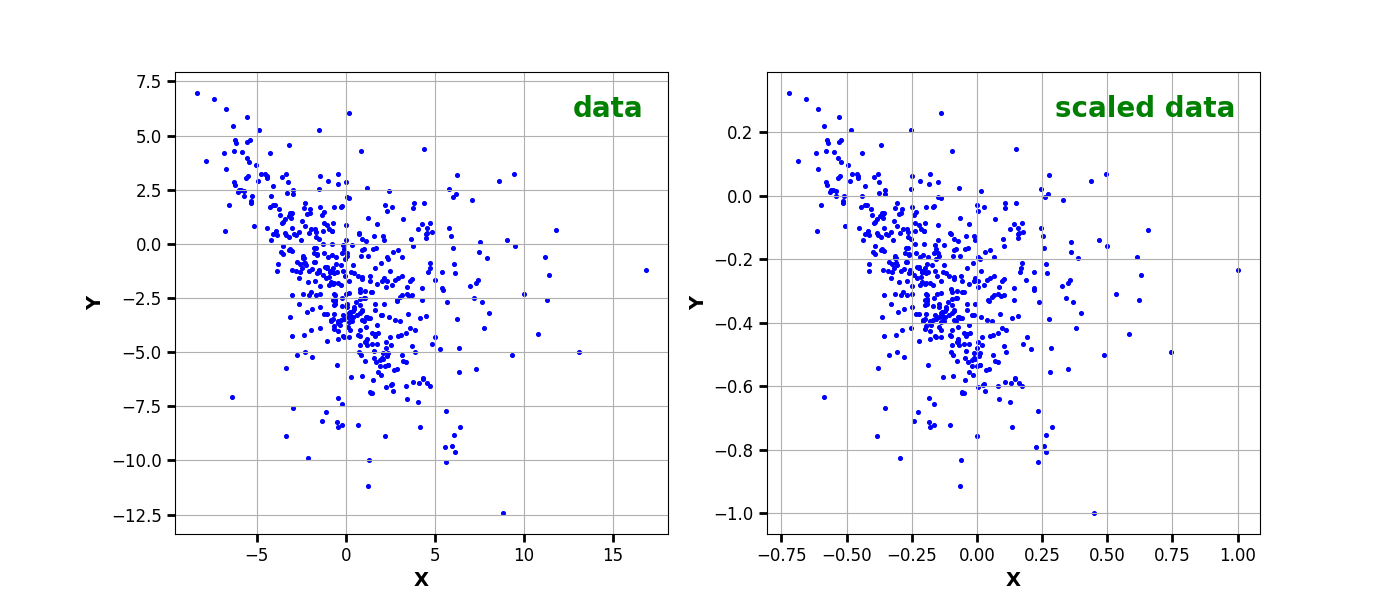} 
  \caption{Data points sampled from one mog distribution with 3 components, MI between X and Y is 0.316.} 
  \label{fig:num_3} 
\end{figure}

\begin{figure}[!h] 
  \centering 
  \includegraphics[width=0.8\textwidth]{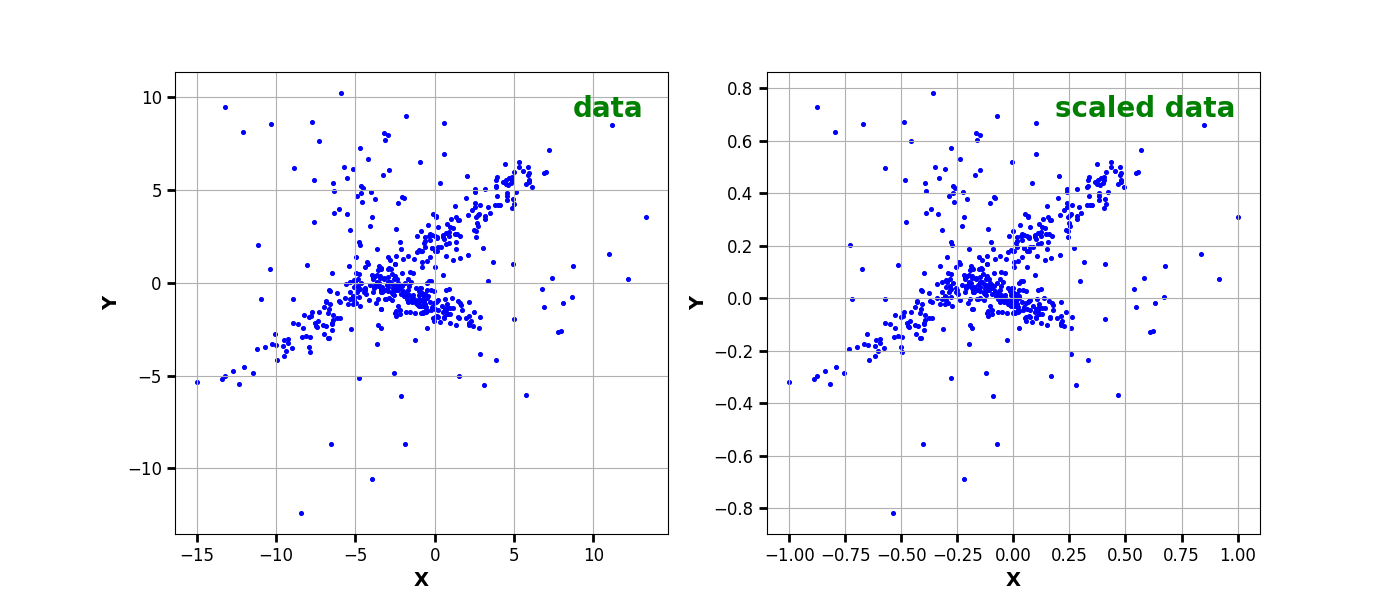} 
  \caption{Data points sampled from one mog distribution with 7 components, MI between X and Y is 0.510. } 
  \label{fig:num_7} 
\end{figure}

\begin{figure}[t] 
  \centering 
  \includegraphics[width=0.8\textwidth]{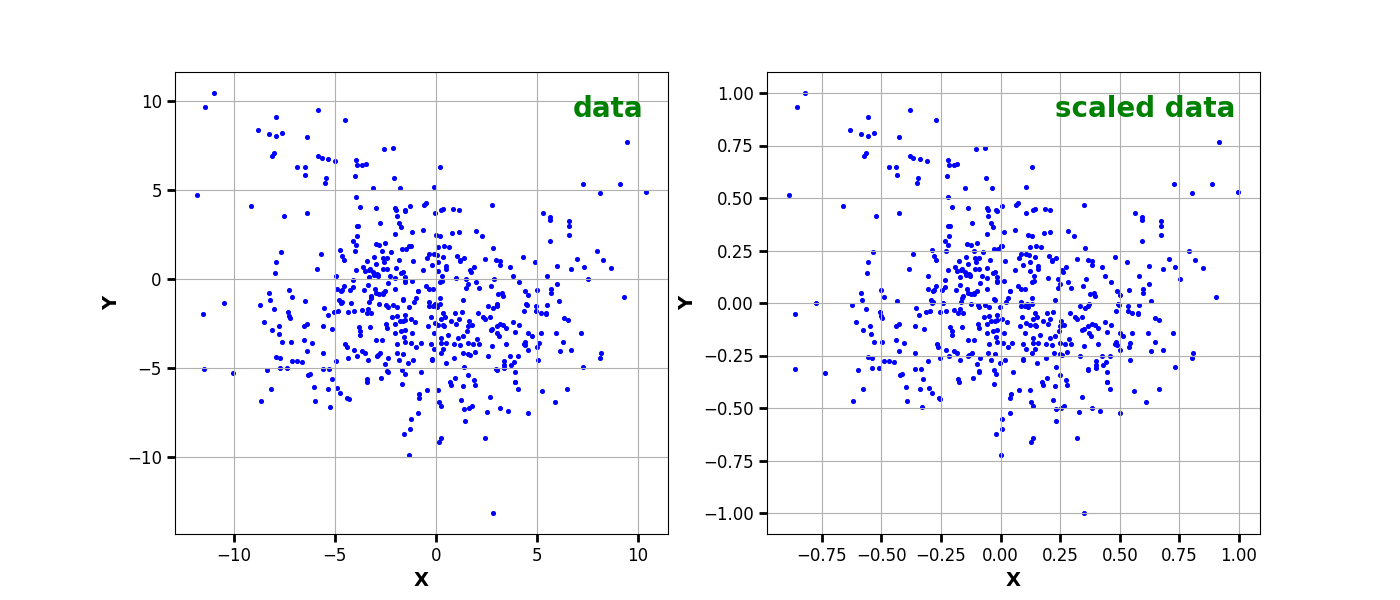} 
  \caption{Data points sampled from one mog distribution with 10 components, MI between X and Y is 0.071. } 
  \label{fig:num_10} 
\end{figure}

\end{document}